\documentclass[%
reprint,
superscriptaddress,
nofootinbib,
amsmath,
amssymb,
aps,
prd,
floatfix,
]{revtex4-2}

\usepackage{xcolor}
\usepackage{xspace}
\usepackage{graphicx}
\usepackage{dcolumn}
\usepackage{bm}
\usepackage{hyperref}
\usepackage[capitalize]{cleveref}
\usepackage[mathlines]{lineno}
\nolinenumbers\relax 
\usepackage{tabularx}
\usepackage{ragged2e}
\usepackage{acronym}
\usepackage{booktabs}
\usepackage{siunitx}

\DeclareSIUnit\year{yr}
\DeclareSIUnit\parsec{pc}
\DeclareSIUnit\msun{\Msun}

\hypersetup{
    colorlinks,
    linkcolor={blue!50!black},
    citecolor={blue!50!black},
    urlcolor={blue!50!black}
}

\begin{document}


\newcommand{\soft}[1]{\texttt{#1}}

\newcommand{\ASTROPY}{\soft{Astropy}\xspace}
\newcommand{\GSTLAL}{\soft{GstLAL}\xspace}
\newcommand{\BAYESTAR}{\soft{BAYESTAR}\xspace}
\newcommand{\CWB}{\soft{cWB}\xspace}
\newcommand{\CWBTWOG}{\soft{cWB-2G}\xspace}
\newcommand{\CWBBBH}{\soft{cWB-BBH}\xspace}
\newcommand{\PYCBC}{\soft{PyCBC}\xspace}
\newcommand{\MBTA}{\soft{MBTA}\xspace}
\newcommand{\SPIIR}{\soft{SPIIR}\xspace}
\newcommand{\LALINFERENCE}{\soft{LALInference}\xspace}
\newcommand{\BAYESWAVE}{\soft{BayesWave}\xspace}
\newcommand{\BILBY}{\soft{Bilby}\xspace}
\newcommand{\GPBILBY}{\soft{GPBilby}\xspace}
\newcommand{\CELERITE}{\soft{celerite}\xspace}
\newcommand{\RIFT}{\soft{RIFT}\xspace}
\newcommand{\LALSUITE}{\soft{LALSuite}\xspace}
\newcommand{\BILBYPIPE}{\soft{BilbyPipe}\xspace}
\newcommand{\PBILBY}{\soft{ParallelBilby}\xspace}
\newcommand{\ASIMOV}{\soft{Asimov}\xspace}
\newcommand{\PESUMMARY}{\soft{PESummary}\xspace}
\newcommand{\PYTHON}{\soft{Python}\xspace}
\newcommand{\NUMPY}{\soft{NumPy}\xspace}
\newcommand{\SCIPY}{\soft{SciPy}\xspace}
\newcommand{\MATPLOTLIB}{\soft{Matplotlib}\xspace}
\newcommand{\SEABORN}{\soft{seaborn}\xspace}
\newcommand{\GWPY}{\soft{GWpy}\xspace}
\newcommand{\DYNESTY}{\soft{Dynesty}\xspace}
\newcommand{\CBCFLOW}{\soft{cbcflow}\xspace}
\newcommand{\GRACEDB}{\soft{GraceDB}\xspace}
\newcommand{\MANIFOLD}{\soft{manifold}\xspace}
\newcommand{\CONDA}{\soft{conda}\xspace}
\newcommand{\LALSIMULATION}{\soft{LALSimulation}\xspace}
\newcommand{\Python}{\soft{Python}\xspace}

\newcommand{\TAYLOR}{\soft{Taylor}\xspace}
\newcommand{\SPINTAYLOR}{\soft{SpinTaylor}\xspace}
\newcommand{\STTFOUR}{\soft{SpinTaylorT4}\xspace}
\newcommand{\TFTWO}{\soft{TaylorF2}\xspace}
\newcommand{\Phenom}{\soft{Phenom}\xspace}
\newcommand{\IMRPhenom}{\soft{IMRPhenom}\xspace}
\newcommand{\IMRPhenomA}{\soft{IMRPhenomA}\xspace}
\newcommand{\IMRPhenomB}{\soft{IMRPhenomB}\xspace}
\newcommand{\IMRPhenomC}{\soft{IMRPhenomC}\xspace}
\newcommand{\IMRPhenomD}{\soft{IMRPhenomD}\xspace}
\newcommand{\IMRPhenomDNRTidal}{\soft{IMRPhenomD\_NRTidal}\xspace}
\newcommand{\IMRPhenomHM}{\soft{IMRPhenomHM}\xspace}
\newcommand{\IMRPhenomP}{\soft{IMRPhenomP}\xspace}
\newcommand{\IMRPhenomPVTWO}{\soft{IMRPhenomPv2}\xspace}
\newcommand{\IMRPhenomPVTHREE}{\soft{IMRPhenomPv3}\xspace}
\newcommand{\IMRPhenomPVTHREEHM}{\soft{IMRPhenomPv3HM}\xspace}
\newcommand{\IMRPhenomXAS}{\soft{IMRPhenomXAS}\xspace}
\newcommand{\IMRPhenomXHM}{\soft{IMRPhenomXHM}\xspace}
\newcommand{\IMRPhenomXP}{\soft{IMRPhenomXP}\xspace}
\newcommand{\IMRPhenomXPHM}{\soft{IMRPhenomXPHM}\xspace}
\newcommand{\IMRPhenomXPHMST}{\soft{IMRPhenomXPHM\_SpinTaylor}\xspace}
\newcommand{\IMRPhenomXOFOURa}{\soft{IMRPhenomXO4a}\xspace}
\newcommand{\EOB}{\soft{EOB}\xspace}
\newcommand{\TEOB}{\soft{TEOB}\xspace}
\newcommand{\TEOBDALI}{\soft{TEOBResumS\_Dali}\xspace}
\newcommand{\SEOBNR}{\soft{SEOBNR}\xspace}
\newcommand{\SEOBNRONE}{\soft{SEOBNRv1}\xspace}
\newcommand{\SEOBNRTWO}{\soft{SEOBNRv2}\xspace}
\newcommand{\SEOBNRTHREE}{\soft{SEOBNRv3}\xspace}
\newcommand{\SEOBNRFOUR}{\soft{SEOBNRv4}\xspace}
\newcommand{\SEOBNRFOUROPT}{\soft{SEOBNRv4\_opt}\xspace}
\newcommand{\SEOBNRFOURROM}{\soft{SEOBNRv4\_ROM}\xspace}
\newcommand{\SEOBNRFOURP}{\soft{SEOBNRv4P}\xspace}
\newcommand{\SEOBNRFOURHM}{\soft{SEOBNRv4HM}\xspace}
\newcommand{\SEOBNRFOURHMROM}{\soft{SEOBNRv4HM\_ROM}\xspace}
\newcommand{\SEOBNRFOUREHM}{\soft{SEOBNRv4EHM}\xspace}
\newcommand{\SEOBNRFOURPHM}{\soft{SEOBNRv4PHM}\xspace}
\newcommand{\SEOBNRFOURT}{\soft{SEOBNRv4T}\xspace}
\newcommand{\SEOBNRFOURTSUR}{\soft{SEOBNRv4T\_surrogate}\xspace}
\newcommand{\SEOBNRFIVE}{\soft{SEOBNRv5}\xspace}
\newcommand{\SEOBNRFIVEP}{\soft{SEOBNRv5P}\xspace}
\newcommand{\SEOBNRFIVEHM}{\soft{SEOBNRv5HM}\xspace}
\newcommand{\SEOBNRFIVEEHM}{\soft{SEOBNRv5EHM}\xspace}
\newcommand{\SEOBNRFIVEROM}{\soft{SEOBNRv5\_ROM}\xspace}
\newcommand{\SEOBNRFIVEPHM}{\soft{SEOBNRv5PHM}\xspace}
\newcommand{\SEOBNRFIVEHMROM}{\soft{SEOBNRv5HM\_ROM}\xspace}
\newcommand{\IMRPhenomPTWONRTidal}{\soft{IMRPhenomPv2\_NRTidal}\xspace}
\newcommand{\IMRPhenomPTWONRTidalTWO}{\soft{IMRPhenomPv2\_NRTidalv2}\xspace}
\newcommand{\IMRPhenomXPNRTidalTWO}{\soft{IMRPhenomXP\_NRTidalv2}\xspace}
\newcommand{\IMRPhenomXPNRTidalTHREE}{\soft{IMRPhenomXP\_NRTidalv3}\xspace}
\newcommand{\IMRPhenomNSBH}{\soft{IMRPhenomNSBH}\xspace}
\newcommand{\SEOBNRFOURNRTidal}{\soft{SEOBNRv4\_ROM\_NRTidal}\xspace}
\newcommand{\SEOBNRFOURNRTidalTWO}{\soft{SEOBNRv4\_ROM\_NRTidalv2}\xspace}
\newcommand{\SEOBNRFOURNRtidalTWONSBH}{\soft{SEOBNRv4\_ROM\_NRTidalv2\_NSBH}\xspace}
\newcommand{\SEOBNRFIVEROMNRTidalTHREE}{\soft{SEOBNRv5\_ROM\_NRTidalv3}\xspace}
\newcommand{\NRSUR}{\soft{NRSurrogate}\xspace}
\newcommand{\SURSEVENDQFOUR}{\soft{NRSur7dq4}\xspace}
\newcommand{\SURSEVENDQTWO}{\soft{NRSur7dq2}\xspace}
\newcommand{\TEOBResumS}{\soft{TEOBResumS}\xspace}

\newcommand{\GWTCTwoPointOne}{GWTC-2.1\xspace}
\newcommand{\GWTCThree}{GWTC-3\xspace}
\newcommand{\GWTCFour}{GWTC-4.0\xspace}

\newcommand{\data}{\ensuremath{\mathbf{d}}\xspace}
\newcommand{\wdata}{\ensuremath{\data_w}\xspace}
\newcommand{\res}{\ensuremath{\mathbf{r}}\xspace}
\newcommand{\model}{\ensuremath{\bm{\mu}}\xspace}
\newcommand{\wmodel}{\ensuremath{\model_w}\xspace}
\newcommand{\parameters}{\ensuremath{\theta}\xspace}
\newcommand{\gpparameters}{\ensuremath{\alpha}\xspace}
\newcommand{\likelihood}{\ensuremath{\mathcal{L}}\xspace}
\newcommand{\prior}{\ensuremath{\pi}\xspace}
\newcommand{\evidence}{\ensuremath{\mathcal{Z}}\xspace}
\newcommand{\BF}{\ensuremath{{\mathcal{B}}}}

\newcommand{\GWFifteen}{\href{https://gwosc.org/eventapi/html/GWTC-2.1-confident/GW150914/v4/}{GW150914}\xspace}
\newcommand{\GWSeventeen}{\href{https://gwosc.org/eventapi/html/GWTC-2.1-confident/GW170814/v4/}{GW170814}\xspace}
\newcommand{\GWNineteen}{\href{https://gwosc.org/eventapi/html/GWTC-3-confident/GW191109_010717/v2/}{GW191109}\xspace}
\newcommand{\GWNineteenLong}{\href{https://gwosc.org/eventapi/html/GWTC-3-confident/GW191109_010717/v2/}{GW191109\_010717}\xspace}
\newcommand{\GWTwentyThreeSix}{\href{https://gwosc.org/eventapi/html/GWTC-4.0/GW230630_070659/v1/}{GW230630}\xspace}
\newcommand{\GWTwentyThreeSixLong}{\href{https://gwosc.org/eventapi/html/GWTC-4.0/GW230630_070659/v1/}{GW230630\_070659}\xspace}
\newcommand{\GWTwentyThreeEight}{\href{https://gwosc.org/eventapi/html/O4_Discovery_Papers/GW230814_230901/v2/}{GW230814}\xspace}
\newcommand{\GWTwentyThreeEightLong}{\href{https://gwosc.org/eventapi/html/O4_Discovery_Papers/GW230814_230901/v2/}{GW230814\_230901}\xspace}
\newcommand{\GWTwentyThreeElevenThirteen}{\href{https://gwosc.org/eventapi/html/GWTC-4.0/GW231113_122623/v1/}{GW231113}\xspace}
\newcommand{\GWTwentyThreeElevenThirteenLong}{\href{https://gwosc.org/eventapi/html/GWTC-4.0/GW231113_122623/v1/}{GW231113\_122623}\xspace}
\newcommand{\GWTwentyThreeElevenTwentyThree}{\href{https://gwosc.org/eventapi/html/GWTC-4.0/GW231123_135430/v2/}{GW231123}\xspace}
\newcommand{\GWTwentyThreeElevenTwentyThreeLong}{\href{https://gwosc.org/eventapi/html/GWTC-4.0/GW231123_135430/v2/}{GW231123\_135430}\xspace}

\newcommand{\numchecked}[1]{\num{#1}}
\newcommand{\numnotchecked}[1]{\textcolor{red}{\num{#1}}}
\newcommand{\qtychecked}[2]{\qty{#1}{#2}}
\newcommand{\qtynotchecked}[2]{\textcolor{red}{\qty{#1}{#2}}}


\newcommand{\Msun}{\ensuremath{\mathit{M_\odot}}}
\newcommand\Mpcyr{\ensuremath{\mathrm{Mpc}^{3}\,\mathrm{yr}}}
\newcommand\Gpcyr{\ensuremath{\mathrm{Gpc}^{3}\,\mathrm{yr}}}
\newcommand\perMpcyr{\ensuremath{\mathrm{Mpc}^{-3}\,\mathrm{yr}^{-1}}}
\newcommand\perGpcyr{\ensuremath{\mathrm{Gpc}^{-3}\,\mathrm{yr}^{-1}}}


\newcommand{\tgeo}{\ensuremath{t_{\text{geo}}}}

\newcommand{\massone}{\ensuremath{m_1}}
\newcommand{\masstwo}{\ensuremath{m_2}}
\newcommand{\Mc}{\ensuremath{\mathcal{M}}}
\newcommand{\Mtot}{\ensuremath{M}}
\newcommand{\Mf}{\ensuremath{M_\mathrm{f}}}
\newcommand{\mratio}{\ensuremath{q}}
\newcommand{\mratiosym}{\ensuremath{\eta}}
\newcommand{\massratio}{\ensuremath{q}}

\newcommand{\mpri}{\massone}
\newcommand{\msec}{\masstwo}
\newcommand{\mchirp}{\Mc}
\newcommand{\mtotal}{\Mtot}

\newcommand{\detectorframe}[1]{\ensuremath{#1^d}}
\newcommand{\mpridet}{\detectorframe{\mpri}}
\newcommand{\msecdet}{\detectorframe{masstwo}}
\newcommand{\mchirpdet}{\detectorframe{\mchirp}}
\newcommand{\mtotaldet}{\detectorframe{\mtotal}}

\newcommand{\Erad}{\ensuremath{E_\mathrm{rad}}}
\newcommand{\lumpeak}{\ensuremath{\ell_{\text{peak}}}}

\newcommand{\energyrad}{\Erad}

\newcommand{\chieff}{\ensuremath{\chi_\mathrm{eff}}}
\newcommand{\chip}{\ensuremath{\chi_\mathrm{p}}}
\newcommand{\chif}{\ensuremath{\chi_\mathrm{f}}}
\newcommand{\spintilt}[1]{\ensuremath{\theta_{#1}}}
\newcommand{\spinone}{\ensuremath{\chi_1}}
\newcommand{\vecspinone}{\ensuremath{\boldsymbol{\chi}_1}}
\newcommand{\spintwo}{\ensuremath{\chi_2}}
\newcommand{\vecspintwo}{\ensuremath{\boldsymbol{\chi}_2}}
\newcommand{\LNewton}{\ensuremath{\hat{\boldsymbol{L}}_\mathrm{N}}}
\newcommand{\thetaJN}{\ensuremath{\theta_{JN}}}

\newcommand{\spinpri}{\spinone}
\newcommand{\spinsec}{\spintwo}
\newcommand{\spineff}{\chieff}
\newcommand{\spinprec}{\chip}

\newcommand{\DA}{\ensuremath{D_\mathrm{A}}}
\newcommand{\DL}{\ensuremath{D_\mathrm{L}}}
\newcommand{\DC}{\ensuremath{D_\mathrm{c}}}
\newcommand{\DM}{\ensuremath{D_{\text{M}}}}
\newcommand{\redshift}{\ensuremath{z}}

\newcommand{\distlum}{\DL}
\newcommand{\distm}{\DM}

\newcommand{\ip}[2]{\ensuremath{\langle #1 | #2 \rangle}}

\newcommand\PEpdfp{\ensuremath{p}}
\newcommand\PEdata{\ensuremath{d}}
\newcommand{\PEparameter}{\ensuremath{\boldsymbol{\theta}}}%
\newcommand{\PEparameterIntrinsic}{\ensuremath{\boldsymbol{\theta}_{\rm int}}}
\newcommand{\PEdataTimeDomain}{\ensuremath{{d}}}%
\newcommand{\PEdataFrequencyDomain}{\ensuremath{{\tilde{d}}}}%
\newcommand{\PEmodelh}{\ensuremath{h_{\mathrm{M}}}}%
\newcommand\PEhyparameter{\ensuremath{\boldsymbol\Lambda}}
\newcommand\PEpdf[2][?]{\ensuremath{\PEpdfp({#2}\ifx#1?\else | {#1}\fi)}} 

\newcommand\PEposterior[1][\PEparameter]{\PEpdf[\PEdata]{#1}}
\newcommand\PElikelihood[1][\PEparameter]{\PEpdf[#1]{\PEdata}}
\newcommand\PEevidence[1][?]{\PEpdf[#1]{\PEdata}}

\newcommand\PEpriorpdfpi{\ensuremath{\pi}} 
\newcommand\PEpdfprior[1]{\ensuremath{\PEpriorpdfpi({#1})}} 
\newcommand\PEprior[1][\PEparameter]{\PEpdfprior{#1}} 
\newcommand\PEpriorpe[1][\PEparameter]{{\let\keepPEpriorpdfpi\PEpriorpdfpi\def\PEpriorpdfpi{\keepPEpriorpdfpi_{\text{PE}}}\PEprior[#1]\let\PEpriorpdfpi\keepPEpriorpdfpi}} 

\newcommand{\flow}{\ensuremath{f_\mathrm{low}}\xspace}
\newcommand{\fhi}{\ensuremath{f_\mathrm{high}}\xspace}
\newcommand{\fsamp}{\ensuremath{f_\mathrm{s}}}
\newcommand{\fNyq}{\ensuremath{f_\mathrm{Nyquist}}}
\newcommand{\alphaRoll}{\ensuremath{\alpha^\mathrm{roll\text{-}off}}}

\newcommand{\PE}[0]{\ac{PE}\xspace}
\newcommand{\PDF}[0]{\ac{PDF}\xspace}
\newcommand{\MCMC}[0]{\ac{MCMC}\xspace}

\newcommand{\HzeroSymbol}{\ensuremath{H_{0}}}
\newcommand{\WmSymbol}{\ensuremath{\Omega_{\mathrm{m}}}}

\newcommand{\PN}[0]{\ac{PN}\xspace}
\newcommand{\BBH}[0]{\ac{BBH}\xspace}
\newcommand{\BNS}[0]{\ac{BNS}\xspace}
\newcommand{\NSBH}[0]{\ac{NSBH}\xspace}
\newcommand{\BH}[0]{\ac{BH}\xspace}
\newcommand{\NR}[0]{\ac{NR}\xspace}
\newcommand{\SNR}[0]{\ac{SNR}\xspace}
\newcommand{\IMR}[0]{\ac{IMR}\xspace}
\newcommand{\GR}[0]{\ac{GR}\xspace}
\newcommand{\PSD}[0]{\ac{PSD}\xspace}
\newcommand{\EOS}[0]{\ac{EOS}\xspace}

\newcommand{\DKLchip}{\ensuremath{D_\mathrm{KL}^{\chi_\mathrm{p}}}}
\newcommand{\DKLchieff}{\ensuremath{D_\mathrm{KL}^{\chi_\mathrm{eff}}}}

\newcommand{\rankstat}{\ensuremath{x}}

\newcommand{\VT}{\ensuremath{\langle VT \rangle}\xspace}
\newcommand{\pastro}{\ensuremath{p_{\mathrm{astro}}}\xspace}
\newcommand{\pterr}{\ensuremath{p_{\mathrm{terr}}}\xspace}
\newcommand{\pbbh}{\ensuremath{p_{\mathrm{BBH}}}\xspace}
\newcommand{\pbns}{\ensuremath{p_{\mathrm{BNS}}}\xspace}
\newcommand{\pnsbh}{\ensuremath{p_{\mathrm{NSBH}}}\xspace}
\newcommand{\comovingv}{\ensuremath{V_\mathrm{c}}}
\newcommand{\comovingvt}{\ensuremath{\langle VT_\mathrm{c} \rangle}}
\newcommand{\injspinmax}{\ensuremath{\chi_\mathrm{max}}}

\newcommand{\printFAR}[2][\day^{-1}]{\ac{FAR} $< \qty{#2}{#1}$}
\newcommand{\printFARper}[2][\day]{\ac{FAR} $<1$ per $\qty{#2}{#1}$}

\newcommand{\changed}[1]{\textcolor{red}{#1}}

\title{
Case studies with GPBilby of glitch-contaminated transient gravitational waves
}


\author{Mattia Emma}
\email{mattia.emma@ligo.org}
\affiliation{Royal Holloway University of London, Egham Hill TW20}
\author{Ann-Kristin Malz}
\affiliation{Royal Holloway University of London, Egham Hill TW20}
\author{Adriana Dias}
\affiliation{Royal Holloway University of London, Egham Hill TW20}
\author{Gregory Ashton}
\affiliation{Royal Holloway University of London, Egham Hill TW20}
\affiliation{Mathematical Sciences, University of Southampton, Southampton SO17 1BJ, United Kingdom}

\date{\today}
\begin{abstract}
In their fourth observing run, the LIGO--Virgo--KAGRA gravitational-wave observatories have found hundreds of new signals, but many are contaminated by non-Gaussian transient noise artefacts known as glitches.
Left unaddressed, glitches can bias parameter inference and lead to misleading astrophysical conclusions.
We present a series of case studies using \GPBILBY, a parameter estimation tool that employs a time-domain likelihood jointly modelling the astrophysical signal with a physical waveform and non-Gaussian noise with a Gaussian process.
We first show that when the detector noise is Gaussian, \GPBILBY produces results consistent with those obtained with the standard Gaussian-noise likelihood, and then consider events affected by non-Gaussian features.
For GW231123, the highest-mass binary black hole candidate observed to date, analyses using \IMRPhenomXPHM reveal coherent residual structure that leads to measurable shifts in inferred source parameters. In contrast, analyses employing \SURSEVENDQFOUR show no significant excess residual power and remain consistent across likelihood choices.
This demonstrates that waveform systematics and flexible noise modelling are intrinsically coupled, as the Gaussian process terms can partially absorb coherent waveform mismatches.
For GW191109, we find that evidence for spin misalignment remains robust despite glitches in both LIGO detectors.
For GW230630\_070659, excluded from GWTC-4.0 owing to poor data quality, we find the data to be consistent with a BBH waveform model, with no additional residual power identified by the Gaussian process component.
Overall, these results highlight how \GPBILBY can be used to perform glitch-robust inference and as a tool to understand waveform modelling systematics.
\end{abstract}

\maketitle

\protected\def\protectedacused{\acused}

\acrodef{LIGO}[LIGO]{Laser Interferometer Gravitational-Wave Observatory}
\acrodef{LHO}[LHO]{\ac{LIGO} Hanford observatory}
\acrodef{LLO}[LLO]{\ac{LIGO} Livingston observatory}
\acrodef{KAGRA}[KAGRA]{KAGRA}\acused{KAGRA}
\acrodef{iKAGRA}[iKAGRA]{initial-phase \ac{KAGRA}}
\acrodef{bKAGRA}[bKAGRA]{baseline-design \ac{KAGRA}}
\acrodef{GEO}[GEO]{GEO\,600 \ac{GW} detector}
\acrodef{aLIGO}{Advanced \ac{LIGO}}
\acrodef{A+}{Advanced+ \ac{LIGO}}
\acrodef{Asharp}[\ensuremath{\text{A}^\sharp}]{\ac{LIGO} \acs{Asharp}}
\acrodef{AdV}{Advanced \acl{Virgo}}
\acrodef{AdV+}{Advanced \acl{Virgo}+}
\acrodef{Virgo}{Virgo}\acused{Virgo}
\acrodef{VirgoNEXT}[Virgo\_nEXT]{Virgo\_nEXT}\acused{VirgoNEXT}

\acrodef{LSC}[LSC]{\acs{LIGO} Scientific Collaboration}
\acrodef{LV}[LV]{\acs{LIGO}--\acs{Virgo} Collaboration\protect\protectedacused{LVC}}
\acrodef{LVC}[LV]{\acs{LIGO}--\acs{Virgo} Collaboration\protect\protectedacused{LV}}
\acrodef{LVK}[LVK]{\acs{LIGO}--Virgo--KAGRA}
\acrodef{IGWN}[IGWN]{International \ac{GWH} Observatory Network}

\acrodef{O1}[O1]{first observing run}
\acrodef{O2}[O2]{second observing run}
\acrodef{O3}[O3]{third observing run}
\acrodef{O3a}[O3a]{first half of the third observing run}
\acrodef{O3b}[O3b]{second half of the third observing run}
\acrodef{O3GK}[O3GK]{observing run}
\acrodef{O4}[O4]{fourth observing run}
\acrodef{O4a}[O4a]{first part of the fourth observing run}
\acrodef{O4b}[O4b]{second part of the fourth observing run}
\acrodef{O4c}[O4c]{third part of the fourth observing run}
\acrodef{O5}[O5]{fifth observing run}

\acrodef{BH}[BH]{black hole}
\acrodef{BBH}[BBH]{binary black hole}
\acrodef{BNS}[BNS]{binary neutron star}
\acrodef{IMBH}[IMBH]{intermediate-mass black hole}
\acrodef{NS}[NS]{neutron star}
\acrodef{BHNS}[BHNS]{black hole--neutron star binaries}
\acrodef{NSBH}[NSBH]{neutron star--black hole binary}
\acrodefplural{NSBH}[NSBH]{neutron star--black hole binaries}
\acrodef{PBH}[PBH]{primordial \ac{BH}}
\acrodef{CBC}[CBC]{compact binary coalescence}

\acrodef{GW}[GW]{gravitational wave\protect\protectedacused{GWH}}
\acrodef{GWH}[GW]{gravitational-wave\protect\protectedacused{GW}}
\acrodef{IFO}[IFO]{interferometer}
\acrodef{SNR}[SNR]{signal-to-noise ratio}
\acrodef{FAR}[FAR]{false alarm rate}
\acrodef{IFAR}[IFAR]{inverse false alarm rate}
\acrodef{FAP}[FAP]{false alarm probability}
\acrodef{PSD}[PSD]{power spectral density}
\acrodef{GR}[GR]{general relativity}
\acrodef{NR}[NR]{numerical relativity}
\acrodef{PN}[PN]{post-Newtonian}
\acrodef{EOB}[EOB]{effective-one-body}
\acrodef{ROM}[ROM]{reduced-order model}
\acrodef{IMR}[IMR]{inspiral--merger--ringdown}
\acrodef{PDF}[PDF]{probability density function}
\acrodef{PE}[PE]{parameter estimation}
\acrodef{CL}[CL]{credible level}
\acrodef{EOS}[EoS]{equation of state}
\acrodef{KLD}[KLD]{Kullback--Leibler divergence}
\acrodef{JSD}[JSD]{Jensen--Shannon divergence}
\acrodef{GCN}[GCN]{general coordinates network}
\acrodef{GWTC}[GWTC]{Gravitational-Wave Transient Catalog}
\acrodef{GWOSC}[GWOSC]{Gravitational Wave Open Science Center}

\acrodef{CWB}[cWB]{coherent WaveBurst}
\acrodef{LAL}[LAL]{\ac{LIGO} algorithm library}

\acrodef{CHRoCC}{central heating radius of curvature correction}
\acrodef{NonSENS}{non-stationary estimation and noise subtraction}

\acrodef{PTA}{Pulsar Timing Array}
\acrodef{GP}{Gaussian process}

\acrodef{MCMC}{Markov chain Monte Carlo}
\acrodef{ESS}{effective sample size}

\acrodef{RS}{rejection sampling}
\acrodef{IS}{importance sampling}
\acrodef{PSIS}{Pareto-smoothed \ac{IS}}
\acrodef{PSRS}{Pareto-smoothed \ac{RS}}

\acrodef{PP}{probability probabilty}
\acrodef{ASD}{amplitude spectral density}
\acrodef{IID}{independent and identically distributed}
\acrodef{KDE}{kernel density estimate}

\acrodef{SHO}{simple harmonic oscillator}
\acrodef{WN}{white noise}

\section{Introduction}
\label{lab: introduction}

The \ac{LVK} Collaborations have recently presented \GWTCFour, the fourth iteration of the Gravitational Wave Transient Catalogue \citep{LIGOScientific:2025hdt, LIGOScientific:2025yae, LIGOScientific:2025slb}.
Analysing data from the \acl{LIGO} \citep[LIGO:][]{LIGOScientific:2014pky}, Virgo \citep{VIRGO:2014yos}, and KAGRA \citep{KAGRA:2020tym} interferometric gravitational-wave detectors, the \ac{LVK} has identified \numchecked{218} signals with a probability of astrophysical origin greater than \numchecked{0.5}.
This rich treasure trove of astrophysical observations is bringing new insights into the dark Universe, such as studies of the mass and spin spectrum \citep{GWTC-4-population} from which we can learn about the deaths of massive stars, measurements of the expansion rate of the Universe \citep{GWTC-4-cosmology}, and tests of the nature of gravity \citep{LIGOScientific:2026qni, LIGOScientific:2026fcf, LIGOScientific:2026wpt}.

However, astrophysical inferences rest on the assumptions made during the individual event analyses: that the background noise is stationary, coloured Gaussian noise \citep{LIGOScientific:2019hgc, GWTC-4-methods}.
If this is not the case, the inferred parameters from individual events can be biased \citep{Powell:2018csz, Macas:2022afm, Udall:2024ovp, Ray:2025rtt, Udall:2025bts} and this may have a potential systematic effect when considering the population as a whole.
To mitigate this, the \ac{LVK} performs detailed studies of all events that undergo parameter estimation \citep{LIGOScientific:2025yae}.
Checks are performed using the strain data and auxiliary channels to understand the data quality and identify any non-Gaussian transient artefacts known as glitches \citep{GWTC-4-methods}.
Of the new signals discovered in \GWTCFour, \numchecked{86} met the criteria for parameter estimation, and of these, \numchecked{44} were found to have non-Gaussian transient artefacts in the data surrounding the event.
We note, however, that for many of these, the observed glitches fall outside the region in which bias is known to occur \citep{Hourihane:2025vxc}.

To mitigate the potential impact glitches may have on parameter estimation, a glitch subtraction approach is typically used \citep[see, e.g.][]{Davis:2022ird, GWTC-4-methods, Macas:2023wiw}, in which the glitch is modelled and subtracted from the data.
However, as described in \citet{Udall:2025bts}, this approach is suboptimal: glitch subtraction always leaves behind residual \ac{SNR}, and subtraction of low-\ac{SNR} glitches can leave residuals that further bias inferences.
Therefore, it is better to perform joint inference of the glitch alongside the astrophysical signal, enabling measurements of the astrophysical source properties marginalised over the uncertainty introduced by the glitch.
Several approaches to joint inference have been developed, such as those using a deterministic glitch model \citep{Udall:2022vkv, Hourihane:2022doe, Ghonge:2023ksb} and those using a data-informed glitch model trained on past observations \citep{Malz:2025xdg}.
Alternative approaches aim to address these biases at the likelihood level, for example through modified noise models such as the hyperbolic likelihood proposed by~\citet{Sasli:2026pds}.
In this work, we continue to develop and validate \GPBILBY \citep{Ashton:2022ztk}, a \ac{GP}-enhanced \citep{Rasmussen2004} approach for parameter estimation of transient gravitational wave signals.

\GPBILBY performs inference on the time series strain data \data after whitening using a given \ac{PSD}.
The core idea is to model the astrophysical signal using a deterministic waveform approximant model with astrophysical source parameters \parameters, and the glitch using a \ac{GP} with an associated set of hyperparameters \gpparameters.
Given a set of source parameters, we define the time-domain residual vector $\res_\parameters \equiv \wdata - \wmodel(\parameters)$ between the whitened data \wdata and the whitened model $\wmodel(\parameters)$. The \ac{GP} log-likelihood is then
\begin{align}
    \ln\likelihood(\data | \parameters, \gpparameters)
    = -\frac{1}{2}\res_\parameters^T\Sigma_\gpparameters^{-1}\res_\parameters
    - \frac{1}{2} \ln \left(\left(2\pi\right)^N \left|\Sigma_\gpparameters\right|\right)\,,
    \label{eqn: likelihood}
\end{align}
where $\Sigma_\alpha$ is the covariance matrix predicted by the \ac{GP} model with hyperparameters $\alpha$, and $|\Sigma_\alpha|$ is its determinant.
In practice, \GPBILBY uses \CELERITE \citep{celerite} which approximates \cref{eqn: likelihood}, enabling rapid evaluation of the likelihood within an order of magnitude of the standard likelihood evaluation time.

The \CELERITE software provides a rich interface for building the \ac{GP} kernel used to predict the covariance matrix in \cref{eqn: likelihood} by combining different \emph{terms}.
We use this interface in the following way.
To model the whitened strain data in the absence of an astrophysical signal or glitch, we find that a single white noise \emph{jitter} term is sufficient; its associated parameter $\sigma$ corresponds to the standard deviation of the whitened strain.
To model white noise alongside a glitch with a single frequency component, we find that the \ac{SHO} kernel term works well.
The \ac{SHO} terms in \CELERITE have three parameters: an angular frequency $\omega_0$, a quality factor $Q$, and an amplitude parameter $S_0$ \citep[see Section 4 of][]{celerite}.
All of these terms are parameterised in terms of their natural logarithm.
We also find this term is effective when the data contains multiple temporally isolated artefacts at the same frequency.
For glitches with multiple frequency components, we use multiple numbered \ac{SHO} terms with order-statistics to circumvent the label-switching degeneracy.

We begin this paper in \cref{sec: updates} by discussing updates to the \GPBILBY software; then in \cref{sec: case_studies} we provide a set of case studies on interesting gravitational-wave observations.
Finally, in \cref{sec: discussion}, we discuss the results and future prospects.
\begin{figure*}
    \centering
    \includegraphics[width=\linewidth]{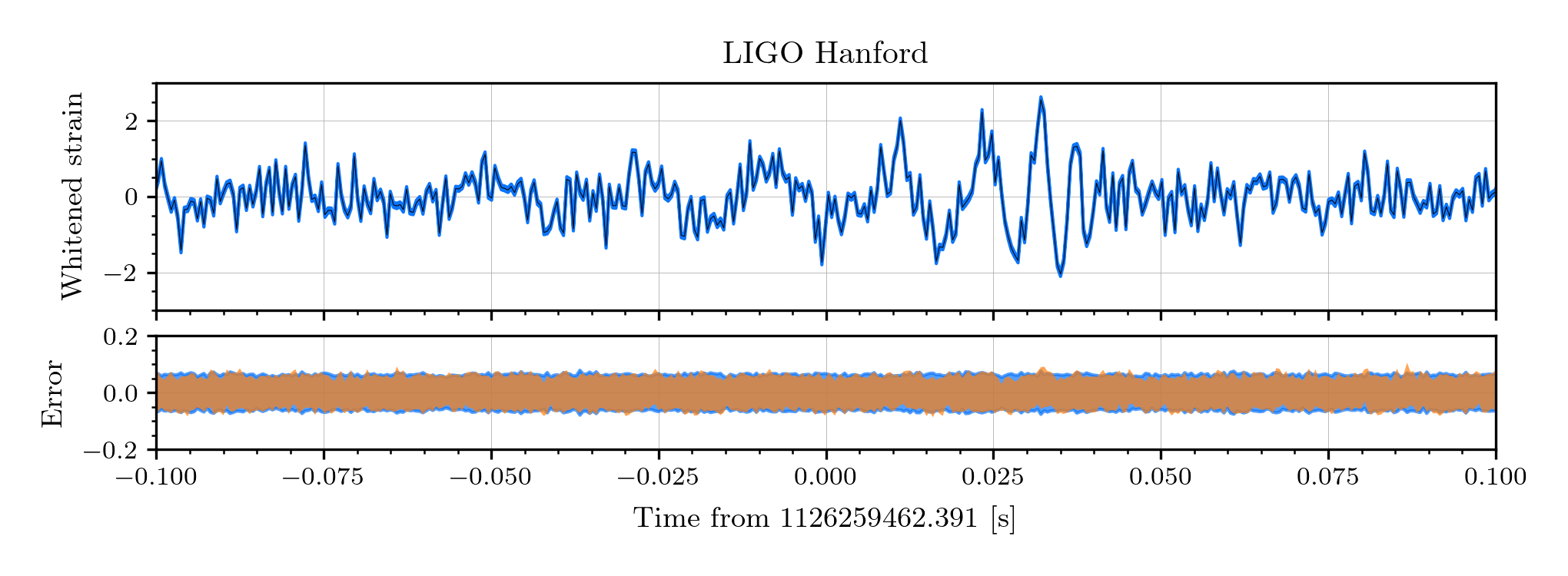}
    \caption{The whitened time-domain strain from the \ac{LHO} data surrounding \GWFifteen.
    In the upper panel, we show the raw whitened strain as a black curve and the measurement errors (estimated from the calibration envelope) as a one-standard-deviation blue band around the whitened strain.
    In the lower panel, we show the measurement error as a blue band, and we show the \qtychecked{68}{\%} non-symmetric interval estimated from the residual directly to validate that the errors are approximately symmetric. Differences in the whitened strain relative to Fig.~6 of~\citet{LIGOScientific:2016vlm} arise from the use of a different \ac{PSD} and data-conditioning procedures.
    }
    \label{fig: calibration}
\end{figure*}

\section{Updates to the software}
\label{sec: updates}
Since the initial version of \GPBILBY \citep{Ashton:2022ztk}, we have improved the software, introducing new \CELERITE terms for \ac{GP} kernel building, developing a new approach to incorporate calibration uncertainty, extending the usability of \GPBILBY to a full range of waveform models, as well as making numerous minor improvements.
Alongside this work, we release Version 0.0.1 of the software \citep{gpbilby2026}.
Below, we describe the first two of these updates in detail with some background on \GPBILBY.

\subsection{Kernel building parameterisation}
\label{sec: parameterisation}
The frequency dependence of the \ac{SHO} kernel term provided by \CELERITE is parameterised by the natural logarithm of the angular frequency $\omega_0$.
However, since the frequency of the \ac{SHO} term corresponds directly to the frequency of the glitch we are trying to model, we have reparameterised the \ac{SHO} term in \GPBILBY in terms of the frequency $f_0=\omega_0/2\pi$.
Beyond the simple parameterisation change, we also use uniform priors on $f_0$ rather than on $\ln\omega_0$; this will provide more prior weight to larger frequencies, but we are typically interested in glitch frequencies that overlap with the signal in the \numchecked{20} to \qtychecked{1000}{Hz} band.
Therefore, we consider a uniform-in-frequency prior appropriate.
Throughout the remainder of this work, we use this new frequency parameterisation of the \ac{SHO} term.

\subsection{Calibration}
\label{sec: calibration}
In the initial implementation of \GPBILBY \citep{Ashton:2022ztk}, the strain errors $\delta h$ provided to \CELERITE were fixed as a user-defined input parameter: for all results in that work, we used a fixed absolute error of \numchecked{0.01}.
In the inference that followed, the \ac{GP} model was flexible enough to model additional error in the kernel by tuning the hyperparameters of the jitter term.
Therefore, in effect, we provided the \ac{GP} with an initial estimate, and then the jitter and \ac{SHO} terms could model the uncertainty.

However, \CELERITE can take as input a vector of heteroscedastic symmetric measurement errors: a 1-standard-deviation error for each data point in the time series.
Therefore, we implemented a new approach to estimate the measurement errors on the whitened strain data from the frequency-domain calibration model.
In standard gravitational-wave parameter estimation, the calibration uncertainty model is constructed as a frequency-domain spline on the amplitude and phase with an associated set of calibration parameters.
Estimates of the uncertainty are obtained from the calibration process as a calibration envelope \citep{Sun:2020wke} which provides the \qtychecked{16}{\%}, \qtychecked{50}{\%}, and \qtychecked{84}{\%} quantiles as a function of frequency.
The envelope is then interpolated with a spline and provided as a prior on the calibration parameters, which are marginalised over during inference \citep{Farr:2014aab, LIGOScientific:2016vlm}.

To approximate the measurement uncertainty on the whitened strain data in \GPBILBY, we take as input the calibration envelope, interpolate over frequency, and fit the quantiles with a skew-normal distribution.
From the fitted skew-normal, we then sample new calibration parameters, use these to perturb the frequency-domain strain, whiten, and Fourier transform, producing many realisations of the whitened time-domain strain, including the effects of calibration uncertainty.
Finally, we take the residuals between the time-domain whitened strain and the perturbed realisations and calculate the standard deviation (as a function of time) across the realisations.
This standard-deviation time series is then provided to \CELERITE as the vector of measurement errors on the whitened strain data.

The \ac{PSD} used for whitening is taken from the \GWTCTwoPointOne re-analysis of the event \citep{LIGOScientific:2021usb}.
In the lower panel of \cref{fig: calibration}, we compare the one-standard-deviation measurement error with the non-symmetric \qtychecked{68}{\%} interval computed across the distribution of residuals. This confirms that our calibration uncertainties are well approximated by a symmetric interval.

\section{Case studies}
\label{sec: case_studies}
To test and validate \GPBILBY, we perform a series of case studies on selected events from the O1-O4a observing runs.
We chose the events \GWFifteen~\cite{LIGOScientific:2016vbw}, \GWSeventeen~\cite{LIGOScientific:2017ycc}, and \GWTwentyThreeEight~\cite{LIGOScientific:2025cmm} as examples where the data is clean, i.e. free from glitches.
We then study \GWNineteen~\cite{Zhang:2023fpp}, \GWTwentyThreeElevenTwentyThree~\cite{LIGOScientific:2025rsn}, and \GWTwentyThreeElevenThirteen to study the performance of our algorithm on events known to be contaminated by glitches.
Finally, we analyse a trigger from the \ac{O4a} which was deemed to be a glitch rather than a gravitational wave event, \GWTwentyThreeSixLong.

\subsection{\GWFifteen}
\label{subsec: GW150914}
\GWFifteen was the first observed \ac{BBH} merger \citep{LIGOScientific:2016aoc} and, with a network \ac{SNR} of \numchecked{24}, remains one of the highest-\ac{SNR} events to date.
Analyses have not identified any data quality issues surrounding the event \citep{LIGOScientific:2016aoc, LIGOScientific:2016vlm, LIGOScientific:2021usb}.
Therefore, as an initial case study, analysing \GWFifteen with \GPBILBY provides a test of the capacity to reproduce standard analysis results when the Gaussian noise assumption holds.

We take as a base analysis the \GWTCTwoPointOne analysis of \GWFifteen \citep{LIGOScientific:2021usb} using the \IMRPhenomXPHM \citep{Pratten:2020ceb} waveform model.
From this, we define six new re-analyses of \GWFifteen, each using \qtychecked{8}{\s} data around the event.
In terms of the likelihood, we use three different configurations: first, an analysis using the standard Whittle likelihood (WL), then a \GPBILBY analysis with a single white noise jitter term (GP-J), and an analysis with a white noise jitter term and a single SHO term (GP-JS).
For each likelihood configuration, we then repeat the analysis using the original \BAYESWAVE \ac{PSD} developed in \GWTCTwoPointOne and an analysis using a new \ac{PSD} created using the version of \BAYESWAVE applied in \GWTCFour \citep{Cornish:2014kda,Littenberg:2014oda,Cornish:2020dwh,Gupta:2023jrn}.
The new \ac{PSD} is available from \citet{Ashton:2025xba}, where it was shown that the differing \ac{PSD} realisation produced small changes in the inferred parameters (consistent with the level expected from \ac{PSD} uncertainty, see \citet{Biscoveanu:2020kat}).
Therefore, the mirrored analysis using different \acp{PSD} provides a point of comparison for comparing the level of difference from the Whittle likelihood and the \GPBILBY analyses to the typical variation due to \ac{PSD} uncertainty.

\begin{figure}
    \centering
    \includegraphics[width=\linewidth]{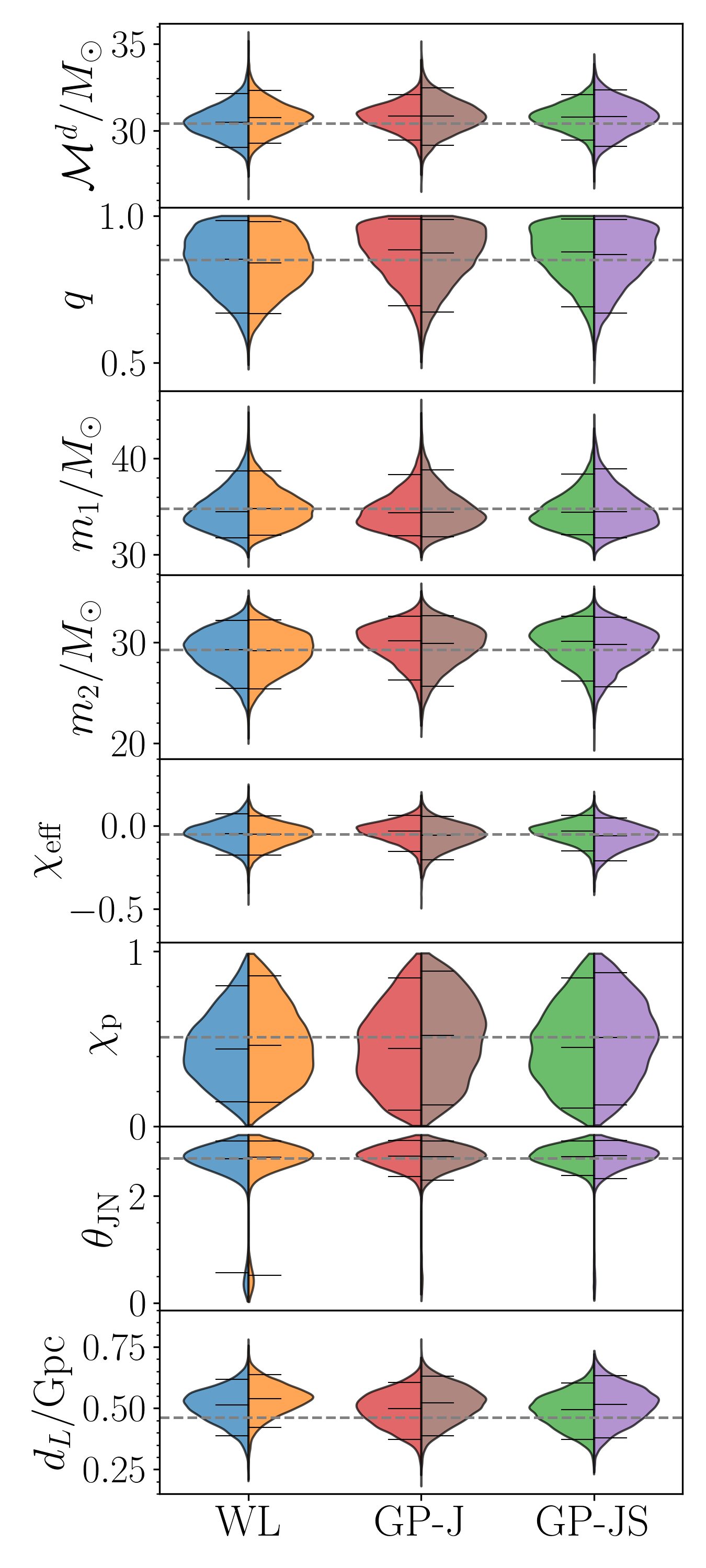}
    \caption{Violin plots of selected \GWFifteen source parameter posteriors for the Whittle likelihood (WL), \GPBILBY with a single jitter term (GP-J), and \GPBILBY with a jitter and SHO term (GP-JS). The left side of each violin shows the posterior obtained using the \GWTCTwoPointOne \ac{PSD}, while the right side shows the posterior obtained using the newly computed \ac{PSD} and the \GWTCFour settings.}
    \label{fig: GW150914_violins}
\end{figure}

In \cref{fig: GW150914_violins}, we show violin plots comparing the one-dimensional posterior distributions across the six analyses for the detector-frame source mass \mchirpdet, mass ratio \mratio, primary mass \mpri, secondary mass \msec, effective inspiral spin \chieff, effective precession \chip, source inclination angle \thetaJN, and luminosity distance \distlum (see Table 3 of \citet{GWTC-4-intro} for definitions and further discussion).
The detector-frame qualifier refers to the mass as measured from the data without including the multiplicative factor $1+z$ required to correct for the redshift $z$ \citep{Krolak:1987ofj}.
Following the \GWTCTwoPointOne methodology, all other masses are given in the source frame and utilise the standard reference cosmology \citep[see][]{GWTC-4-intro}, namely column TT+lowP+lensing+ext of Table 4 in \citet{Planck:2015fie}.

Broadly, \cref{fig: GW150914_violins} shows agreement between the posterior distributions across all configurations of the likelihood.
There are mostly per cent-level differences in the inferred quantiles for most parameters, which are at a similar level to the differences between estimates of the \ac{PSD}.
The exception is the inclination angle \thetaJN, where the Whittle likelihood analyses show more weight in a secondary mode than is measured from any of the \GPBILBY analyses.
We will comment on this further in \cref{sec: discussion}.

\begin{figure}
    \centering
    \includegraphics[width=0.44\linewidth]{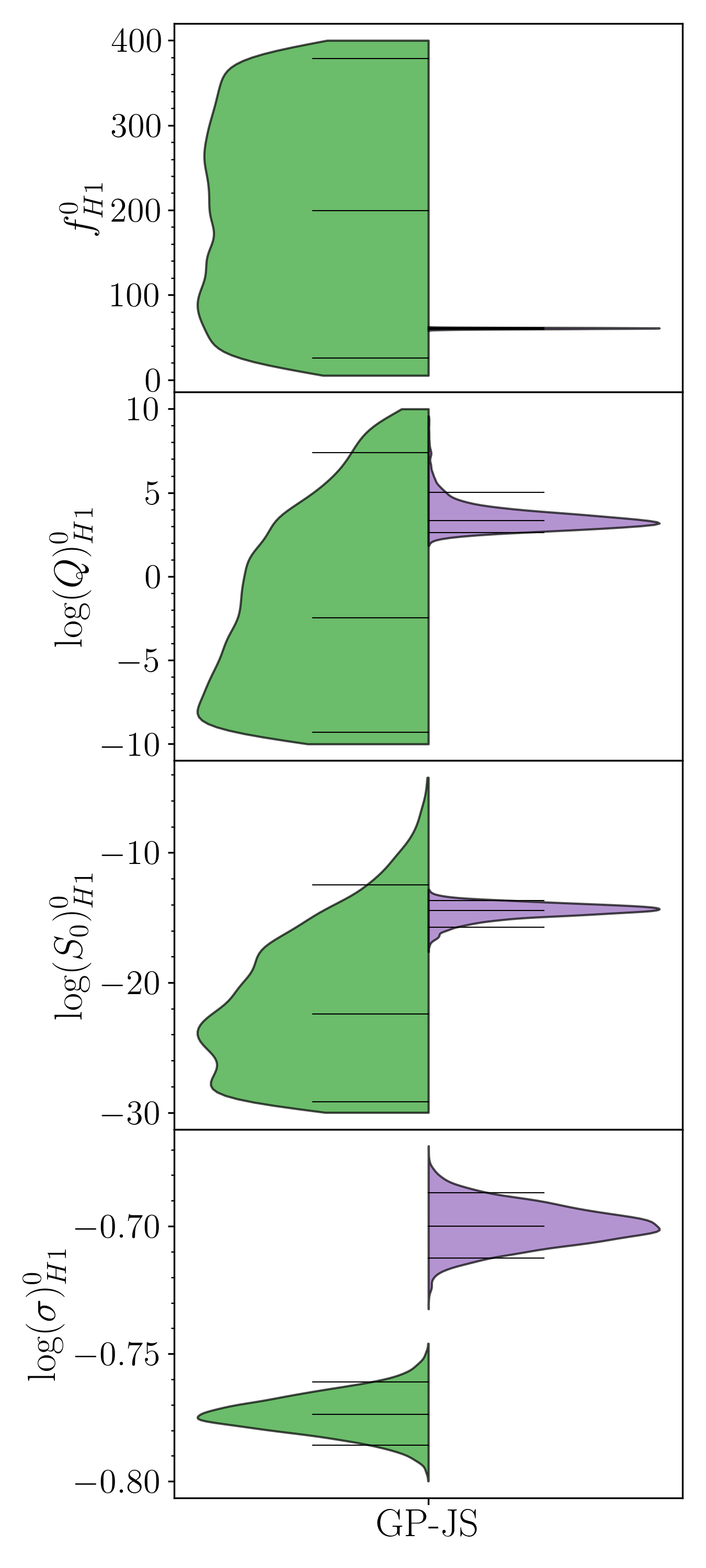}
    \includegraphics[width=0.44\linewidth]{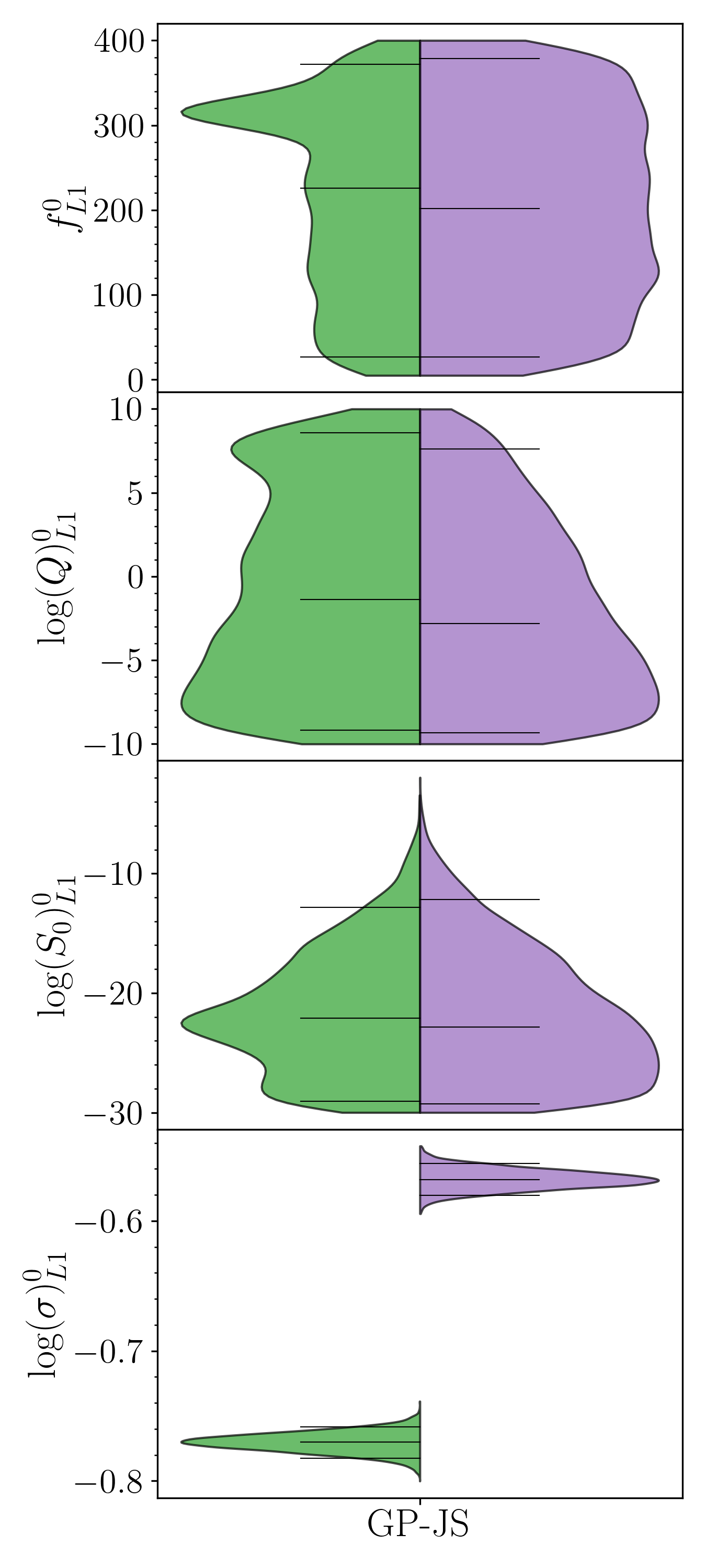}
    \caption{Posterior distributions for the SHO term parameters of the \GPBILBY GP-JS analyses of \GWFifteen (left panel: \ac{LHO}, right panel: \ac{LLO}).
    The analyses using the \GWTCTwoPointOne \ac{PSD} are shown in green, while those using the newly computed \ac{PSD} are shown in purple.
    For the \ac{LHO} analysis with the newly computed \ac{PSD}, the frequency posterior is tightly peaked at $f^{0}_{\rm H1} = 60.53^{+0.98}_{-1.17}$~Hz.
}
    \label{fig: GW150914_GP}
\end{figure}

To investigate the inferences from the \ac{SHO} term in the two GP-JS analyses, in \cref{fig: GW150914_GP}, we provide the posterior distribution for SHO parameters.
Interestingly, we find a spike in the \ac{LHO} (\ac{LLO}) analysis at $\approx$\qtychecked{60}{\Hz} ($\approx$\qtychecked{320}{\Hz}) for the \GWTCTwoPointOne \ac{PSD} (new \ac{PSD}).
That these frequencies lie close to known lines in the strain data \citep{LIGOScientific:2019hgc}, caused by the \qtychecked{60}{\Hz} power line and its harmonics, suggests that the \ac{PSD} has not perfectly whitened the data, and \GPBILBY can find and model these features.
That the \ac{PSD} is unable to model the power lines is not surprising, since it is a coherent deterministic signal with a narrowband: the coloured Gaussian noise assumption is weak under these conditions.
However, standard intuition within the field argues that, since the \ac{PSD} is large close to the lines, these regions have a minimal impact on the likelihood due to the noise-weighting within the Whittle likelihood.

\begin{figure}
    \centering
    \includegraphics[width=\linewidth]{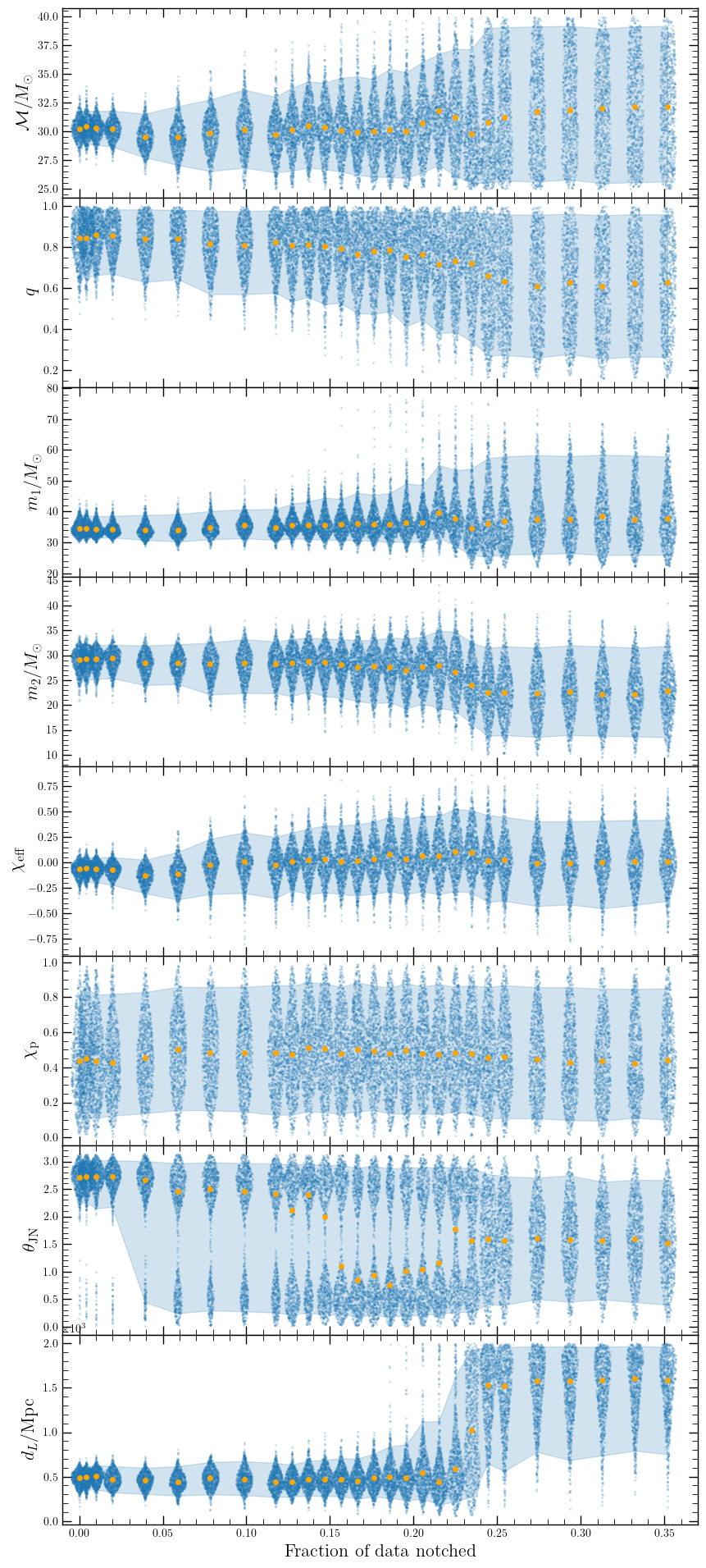}
    \caption{Plots of selected \GWFifteen source parameter posteriors, with respect to the fraction of data notched.
    The blue filled regions show how the \SI{90}{\%} interval changes as the proportion of notched data increases. The orange dots represent the median values and the blue shaded scatter show the full distribution of posterior samples.}
    \label{fig: notching}
\end{figure}

To validate that the power line does not have any impact on the inference of the astrophysical source parameters, we perform a \emph{notching} study.
We take the H1 and L1 \ac{PSD}s, identify frequency-domain windows ranging from \qtychecked{4}{\hertz} to \qtychecked{600}{\hertz} widths around the \qtychecked{60}{\hertz} line and replace the \ac{PSD} within these windows to a fixed value of \numchecked{1}.
We then repeat the analysis under the Whittle likelihood, varying the size of the windows and estimating the posterior distribution. 
In \cref{fig: notching}, we plot the posterior distribution for a set of selected source parameters as a function of the fraction of data removed.
In the extremes, we find that when a negligible amount of data is removed, the posterior inference is robust, while when a significant amount is removed, the posteriors widen and become uninformative as expected.
Between these extremes, we do not find any region in which the posterior is inconsistent with the inferences without notching.
This validates that, while the \ac{PSD} is unable to model the contributions from the power lines, their impact on the inferred posteriors is negligible, agreeing with the standard intuition within the field.

In summary, the case study of \GWFifteen further validates the absence of any transient noise artefacts (at least that \GPBILBY can find and model) impacting the measured source parameters.

\subsection{\GWSeventeen}
\label{subsec: GW170814}
\GWSeventeen is the first gravitational wave event detected by three detectors, namely the two LIGO and the Virgo observatories \citep{LIGOScientific:2017ycc}.
It has a network \ac{SNR} of \numchecked{18}, which, combined with the triple detector observation, enabled a tight sky-localization area of \qtychecked{60}{\deg^2} (\qtychecked{90}{\%} credible region).
Similarly to the approach taken for \GWFifteen, we take as a reference point the \GWTCTwoPointOne~\citep{LIGOScientific:2021usb} analysis and perform three re-analyses using \qtychecked{4}{\second} of data around the event. We perform a WL analysis, a GP-J and a GP-JS analysis, keeping all the other settings fixed.

\begin{figure}[htp!]
    \centering
    \includegraphics[width=\linewidth]{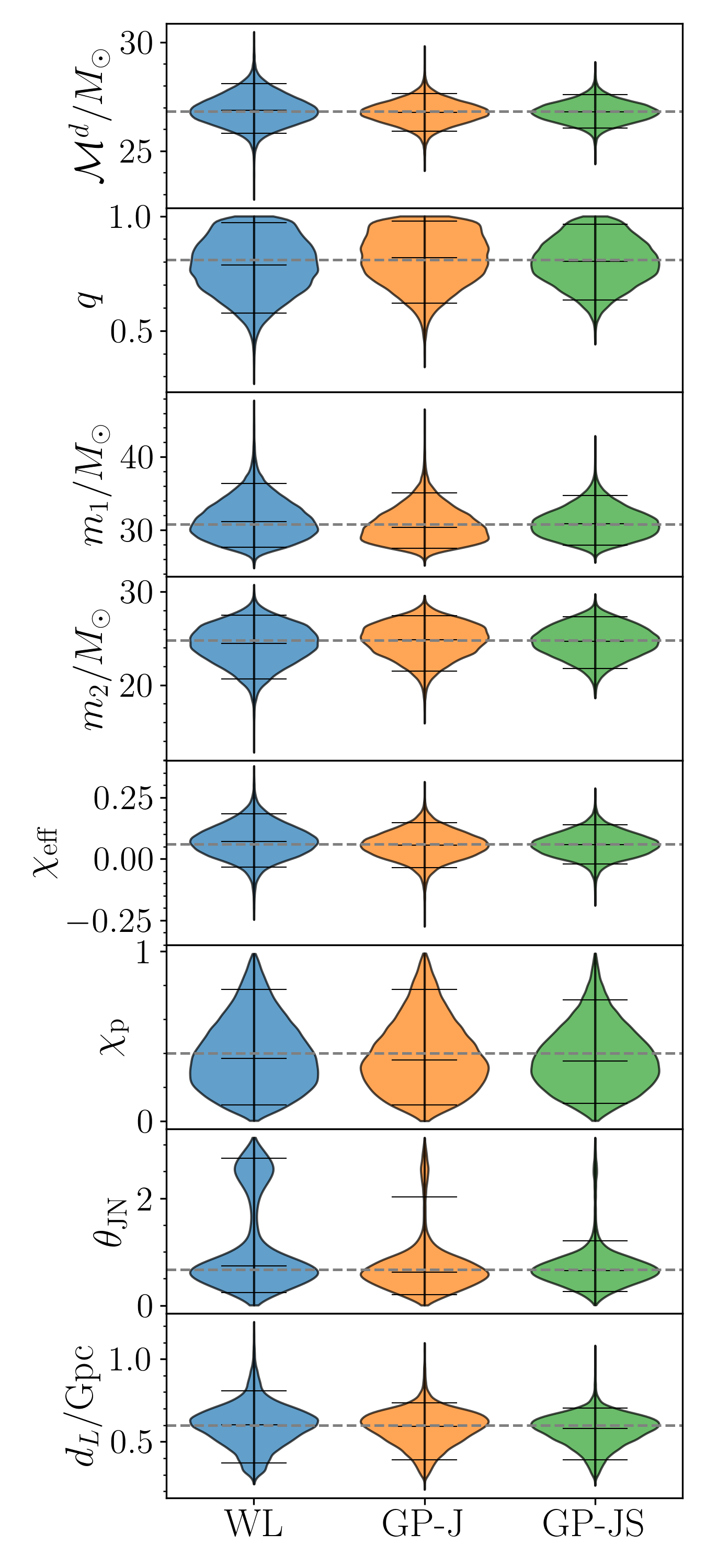}
    \caption{Violin plots of selected \GWSeventeen source parameter posteriors for the Whittle likelihood (WL), \GPBILBY with a single jitter term (GP-J), and \GPBILBY with a jitter and SHO term (GP-JS).}
    \label{fig: GW170814_violins}
\end{figure}

In \cref{fig: GW170814_violins}, we compare the one-dimensional posterior distributions for the three analyses.
The results show a broad agreement between the posteriors in the different configurations, although there is a general tightening of the \qtychecked{90}{\%} credible intervals from the WL to the GP-JS analyses.
As for \GWFifteen, the posterior for the inclination angle \thetaJN~shows more weight for a secondary mode in the WL configuration than in either of the GP analyses.
This support is partially there in the GP-J configuration, but vanishes in the GP-JS analysis.

Considering the \ac{GP} terms in the GP-JS analysis, we find that the posterior on the frequency parameter for both detectors is consistent with the prior, suggesting no excess power is found.
Similarly, both the quality and amplitude parameters rail against the lower prior boundary, confirming that there are no significant non-Gaussian features in those data segments.
Finally, the variance of the white noise term peaks at $\simeq 0.78$ for both detectors, as for the \GWTCTwoPointOne PSD analysis of \GWFifteen.

\subsection{\GWTwentyThreeEight}
\label{sec: GW230814}

\begin{figure}[htp!]
    \centering
    \includegraphics[width=\linewidth]{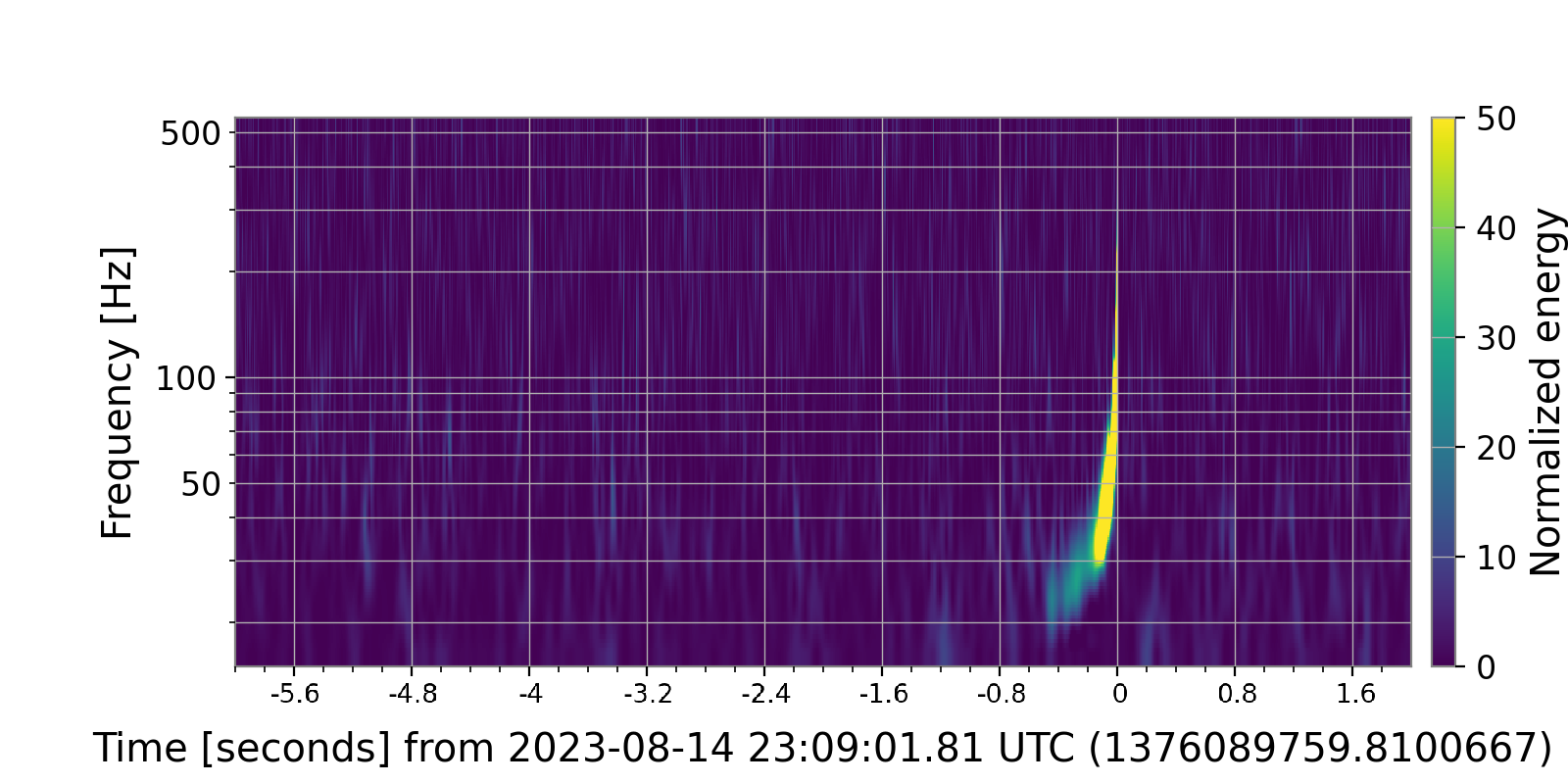}
    \caption{The time-frequency spectrogram showing \ac{LLO} data surrounding \GWTwentyThreeEight.}
    \label{fig: GW230814_qscan}
\end{figure}

To further test our algorithm, we analyze the single-detector event \GWTwentyThreeEightLong (hereafter \GWTwentyThreeEight) from O4a~\cite{LIGOScientific:2025cmm}.
This event has the second loudest network-SNR observed in the current network of ground-based gravitational wave detectors \citep{GWTC-4-results}.
The LVK analysis found no data quality issues \citep{LIGOScientific:2025cmm}, making this an ideal candidate to probe the performance of \GPBILBY in the high-SNR regime with a clean signal (cf. ~\cref{fig: GW230814_qscan}).

We perform a WL, a GP-J and a GP-JS analysis using the settings employed in the \IMRPhenomXPHMST analysis from \citet{LIGOScientific:2025rsn}.
For each analysis, we repeat it with a \qtychecked{4}{\second} and \qtychecked{8}{\second} data duration, and use in both cases an end time of \qtychecked{2}{\second} after the trigger time.

\begin{figure}[htp!]
    \centering
    \includegraphics[width=\linewidth]{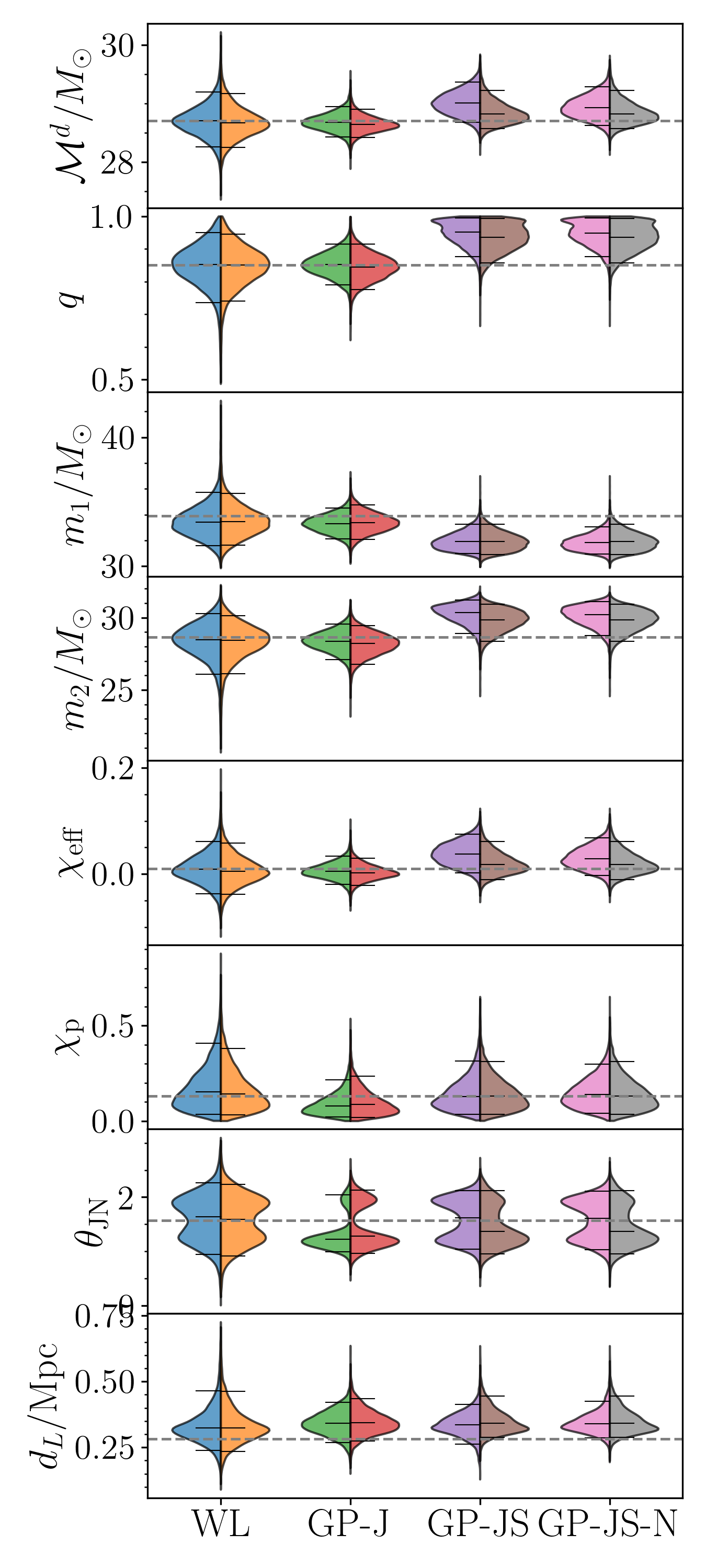}
    \caption{Violin plots showing selected source-parameter posteriors for \GWTwentyThreeEight obtained with different analyses.
For each violin, the left half corresponds to the posterior derived using 4\,s of data, while the right half shows the posterior obtained using 8\,s of data.
From left to right, the violins correspond to analyses performed with \BILBY using the Whittle likelihood, \GPBILBY including only a jitter term, \GPBILBY including both jitter and SHO terms, and \GPBILBY including both jitter and SHO terms applied to data in which a \qtychecked{30}{\hertz} band around \qtychecked{345}{\hertz} has been notched, as discussed in the text.}
    \label{fig: GW230814_posteriors}
\end{figure}

The inference results for selected parameters are shown in \cref{fig: GW230814_posteriors}.
The \BILBY analyses using \qtychecked{4}{\second} and \qtychecked{8}{\second} of data yield consistent results, and these in turn are consistent with the jitter-only \GPBILBY analysis (GP-J), although the credible intervals are narrower.

However, the \GPBILBY analyses that include an SHO term (GP-JS) recover markedly different posterior distributions for several intrinsic parameters.
Specifically, we observe: (i) a slightly higher chirp mass, driven by a lower primary mass and higher secondary mass, (ii) a preference for more equal-mass systems (higher mass ratio), and (iii) a shift in the effective spin parameter $\chi_{\rm eff}$ toward positive values, away from zero.
Notably, while the chirp mass median from the SHO analysis remains within the \qtychecked{90}{\percent} bounds of the Whittle likelihood results, the posteriors show narrower credible intervals.
Furthermore, the \qtychecked{4}{\second} and \qtychecked{8}{\second} SHO analyses show some differences: the \qtychecked{8}{\second} analysis yields chirp mass and effective spin posteriors more similar to the \BILBY results than the \qtychecked{4}{\second} SHO analysis.

\begin{figure}[htp!]
    \centering
    \includegraphics[width=\linewidth]{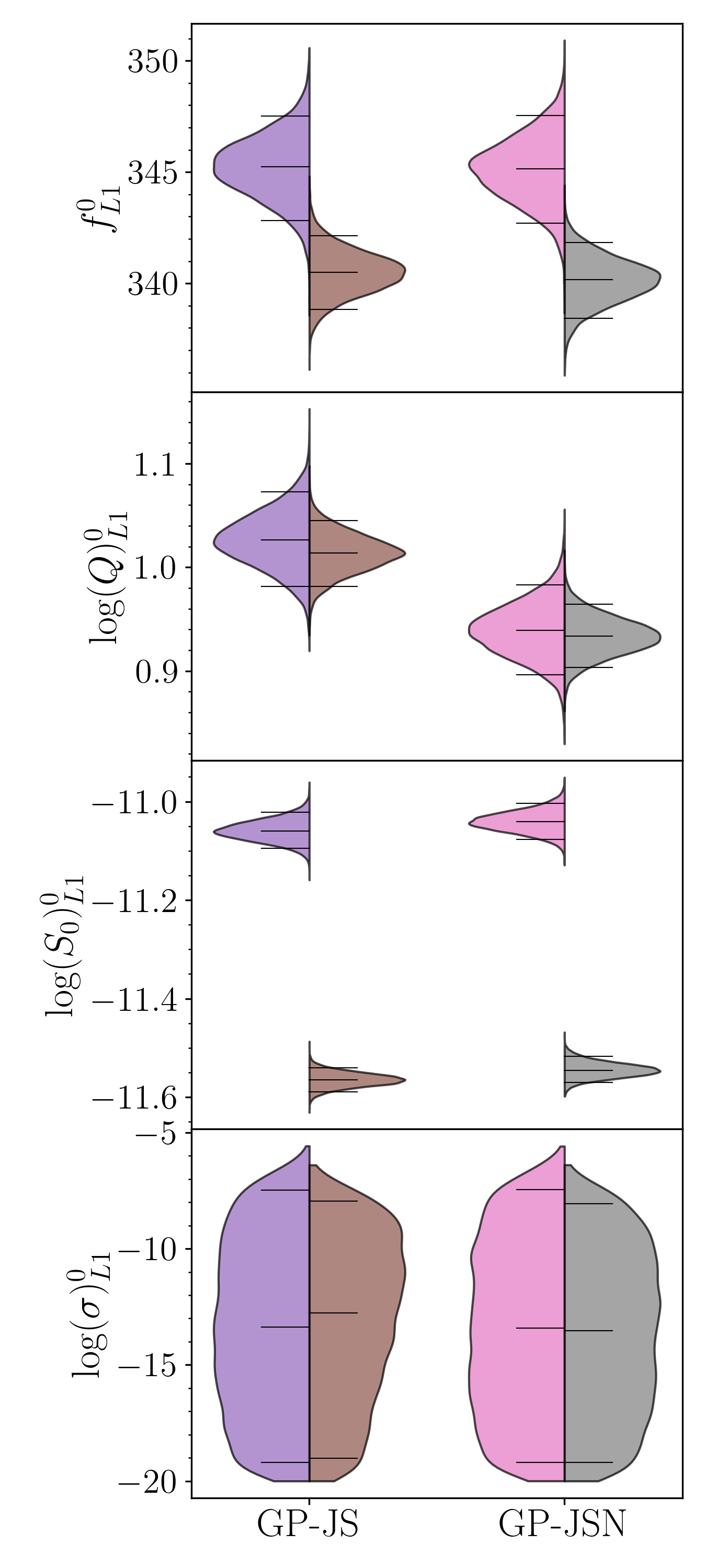}
    \caption{Posterior distributions for the SHO term parameters of the \GPBILBY GP-JS analyses of \GWTwentyThreeEight (\ac{LLO} detector).}
    \label{fig: GW230814_posteriors_SHO}
\end{figure}

Examining the inferred values of the SHO terms in the GP-JS analysis, \cref{fig: GW230814_posteriors_SHO} demonstrates that the SHO term identifies power at $\sim$\qtychecked{345}{\hertz} and $\sim$\qtychecked{340}{\hertz} for the \qtychecked{4}{\second} and \qtychecked{8}{\second} analyses, respectively.
Examination of the time-frequency spectrogram, included in the data release~\cite{emma2026gpbilby_zenodo} and shown in~\citet{LIGOScientific:2025cmm}, reveals no clearly visible power at these frequencies, nor do spectral lines appear in the \ac{PSD} in this frequency range.
Interestingly, the jitter term is unconstrained for this event, signalling difficulties in modelling the distribution of the Gaussian and stationary part of the noise.

To investigate whether the inference changes are driven by the modeling of this $\sim$\qtychecked{345}{\hertz} power as a glitch, we perform additional GP-JS runs using a PSD notched in the frequency range \qtyrange{330}{360}{\hertz}, following the procedure described in Section~\ref{subsec: GW150914} (we designate these notched analyses as GP-JSN).
The posterior distributions for these notched runs, shown in the rightmost violins of \cref{fig: GW230814_posteriors}, agree with those obtained using unnotched data (i.e. GP-JS) for both the \qtychecked{4}{\second} and \qtychecked{8}{\second} data sets.
This suggests that the $\sim$\qtychecked{345}{\hertz} power is not responsible for the differences between the GP-JS analysis and the GP-J and WL analyses.

In \cref{fig: GW230814_posteriors_SHO}, we also include the SHO posterior distributions from the GP-JSN analysis.
Interestingly, the SHO term still identifies power at \qtyrange{340}{345}{\hertz}.
This persistence can be attributed to the fact that notching is performed in the frequency domain, while the data analysis is conducted in the time domain.
The inverse Fourier transform necessarily reintroduces some power to the notched frequency bins, allowing the SHO term to continue modeling features in this frequency range.

\begin{figure*}
    \centering
    \includegraphics[width=\textwidth]{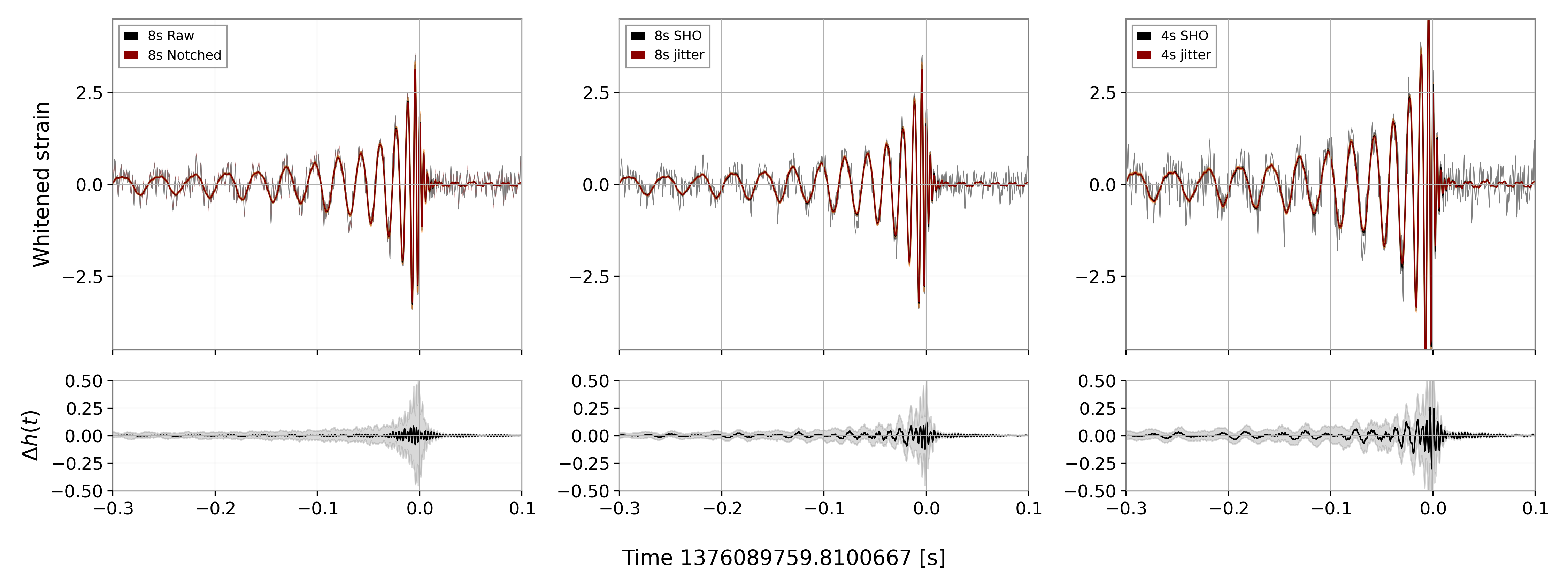}
    \caption{Posterior predictive plots for the event \GWTwentyThreeEight for the \ac{LLO} detector. In each panel we compare the strain prediction given by the maximum likelihood waveform for two analyses using the \IMRPhenomXPHMST waveform.
    The upper plots show the whitened strain data and the predicted strains with 90\% credibility intervals, while the bottom plots show the residuals and their 90 \% credibility intervals. 
    In the left panel we compare two GP-JS analyses, one using 8 s of the full raw data, and one using 8s of data notched between \qtyrange{330}{360}{\hertz}.
    The middle panel compares the 8s GP-JS and GP-J analyses. 
    The right panel compares the 4s GP-JS and GP-J analyses.
    Differences in the whitened strain relative to Fig.~1 of~\citet{LIGOScientific:2025cmm} arise from the use of different data-conditioning procedures.}
    \label{fig:GW230814_predictive_plot}
\end{figure*}

To further investigate the discrepancies between the GP-JS and GP-J analyses, we compare the signal and noise predictions from the GP model across different runs.
\cref{fig:GW230814_predictive_plot} summarizes the comparison between the different \GPBILBY analyses.
The left panel, comparing the 8\,s SHO runs with notched and unnotched data, shows that the main differences arise from the notching procedure, appearing as high-frequency beating patterns that are most evident during and after the merger phase. 
Additional diagnostics based on time--frequency spectrogram of the residuals show that their dominant frequency content coincides with the high-frequency part of the signal, where the largest residual amplitudes are observed.
The comparisons between the SHO and jitter-only runs, shown for both the 4\,s (right panel) and 8\,s analyses (middle panel), reveal that the largest discrepancies occur in the pre-merger and merger stages, where the two models predict different signal morphologies.
These differences are more pronounced in the 4\,s analysis, which exhibits larger-amplitude residuals, consistent with the stronger shifts observed in the inferred source parameters.

Overall, these results indicate that the inference differences between the SHO and jitter-only analyses are primarily driven by the modeling of the pre-merger and merger portions of the signal.
In the high-SNR regime, subtle structures or fluctuations in these phases can be absorbed either by the signal model or by the noise model, depending on the chosen GP kernel.
The SHO term, owing to its ability to capture narrow-band features, absorbs power not accounted for by the jitter-only model, leading to compensatory changes in the inferred source parameters to maintain agreement with the data.

\subsection{\GWNineteen}
\label{subsec: GW191109}

After validating \GPBILBY's ability to accurately infer source parameters for gravitational-wave signals unaffected by nearby glitches, we now turn to events with data quality issues, starting with \GWNineteenLong (hereafter \GWNineteen).
This event is of particular astrophysical interest, as it exhibits strong support for an effective inspiral spin that is anti-aligned with the orbital angular momentum, favouring a dynamical formation scenario for the binary system~\cite{Zhang:2023fpp}.
\GWNineteen is also interesting from the perspective of tests of fundamental physics: consistency tests performed as part of the \GWTCThree analysis identified anomalous behaviour for this event \citep{LIGOScientific:2021sio}, which was subsequently attributed to the presence of glitches in both detectors. 
These features make \GWNineteen a compelling case study for assessing the robustness of parameter inference and tests of general relativity in the presence of non-Gaussian noise.

\subsubsection{Data description: raw and deglitched frames}

Glitches were identified in both the \ac{LHO} (H1) and \ac{LLO} (L1) detectors around the time of \GWNineteen.
In \ac{LHO}, a glitch is present approximately \qtyrange{1.2}{2}{\second} before merger at a frequency of $\sim\qtychecked{36}{\hertz}$.
This transient lies outside the signal track and is therefore not expected to significantly affect source parameter inference.

In \ac{LLO}, the noise is more complex.
The data contain a series of slow-scattering arches \cite{LIGO:2020zwl} that repeat in time and frequency. 
The strongest of these occurs approximately \qtychecked{4}{\second} before the trigger time and is followed by additional arches near \qtychecked{24}{\hertz} and \qtychecked{36}{\hertz}, both of which overlap in time with the gravitational-wave signal track. 
The \ac{LHO} glitch and the \ac{LLO} feature at \qtychecked{24}{\hertz} were subtracted in the deglitched data products used for the \GWTCThree parameter estimation \cite{KAGRA:2021vkt}.
In contrast, the excess power at \qtychecked{36}{\hertz} in \ac{LLO} was not removed, as it could not be unambiguously separated from the signal.

\cref{fig: GW191109qscan}, in \cref{app: spectrograms}, shows time--frequency representations of the raw data, the reconstructed glitches, and the glitch-subtracted frames for both detectors. 
In the following, we analyse both the raw and deglitched data in order to assess the impact of glitch modelling on parameter inference.

\subsubsection{Joint signal--glitch inference with \GPBILBY}

\begin{figure*}[htp!] 
\centering 
\includegraphics[width=8cm]{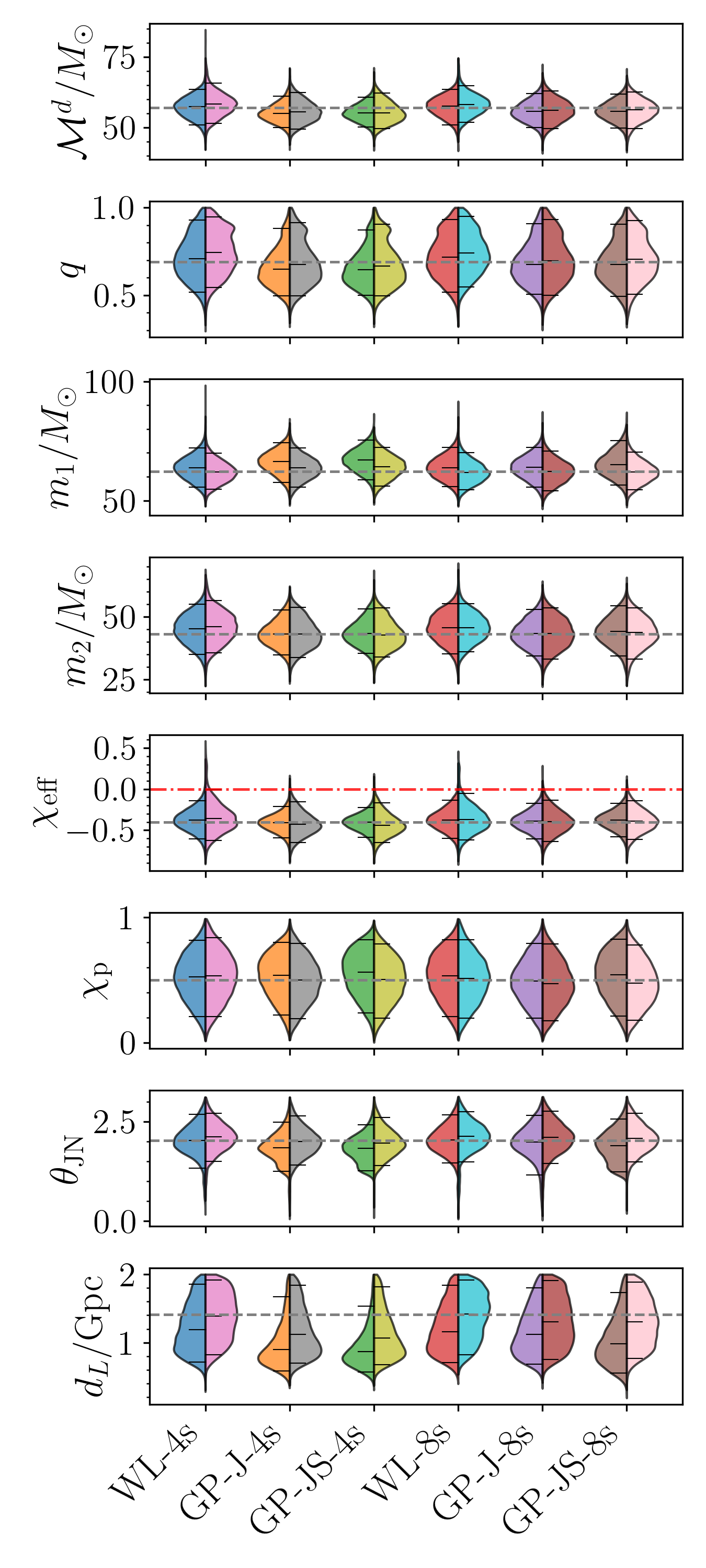} 
\includegraphics[width=8cm]{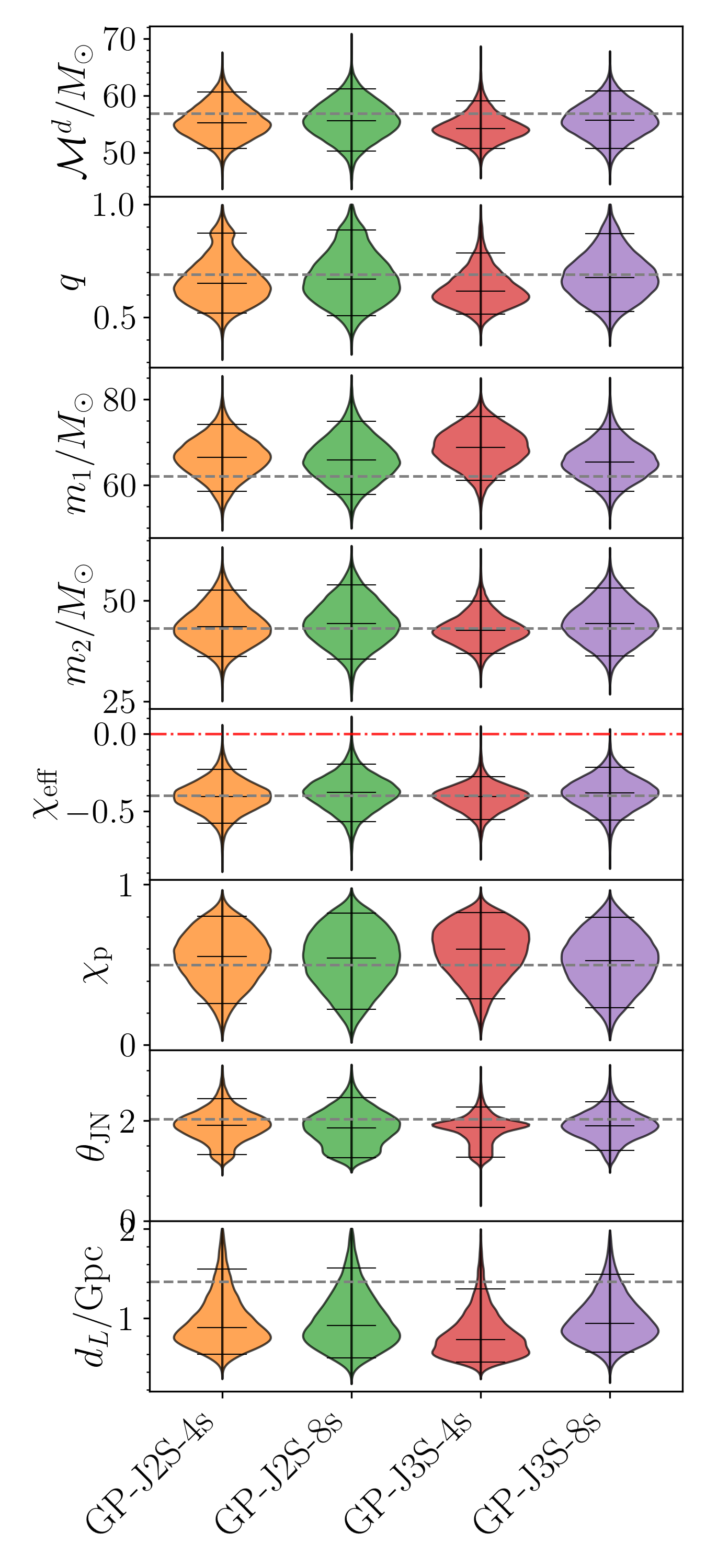} 
\caption{Violin plots of selected \GWNineteenLong source parameter posteriors comparing different analysis configurations. The left panel shows results from \qtychecked{8}{\second} analyses including one SHO term; the right panel shows results with multiple SHO terms using raw data only. WL denotes the Whittle likelihood (\BILBY), GP-J uses only a jitter term, GP-JS uses a jitter and one SHO term, GP-J2S uses two SHO terms, and GP-J3S uses three SHO terms. Within each violin, the left side shows the posterior obtained from the raw data, while the right side shows the posterior obtained from the deglitched data.}
\label{fig: GW191109_8s_4s_posteriors} 
\end{figure*}

\begin{figure*}[htp!] 
\centering 
\includegraphics[width=8cm]{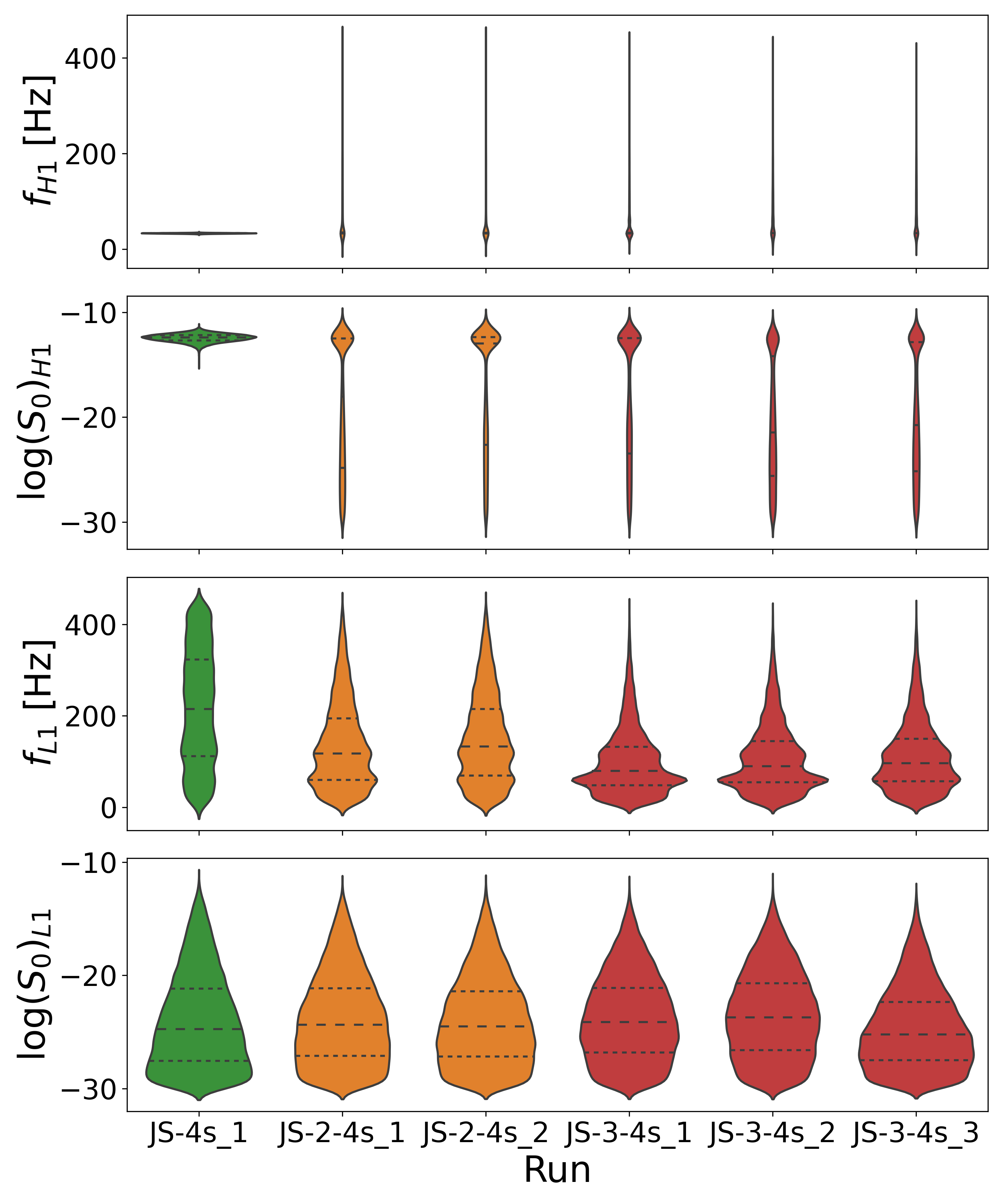} 
\includegraphics[width=8cm]{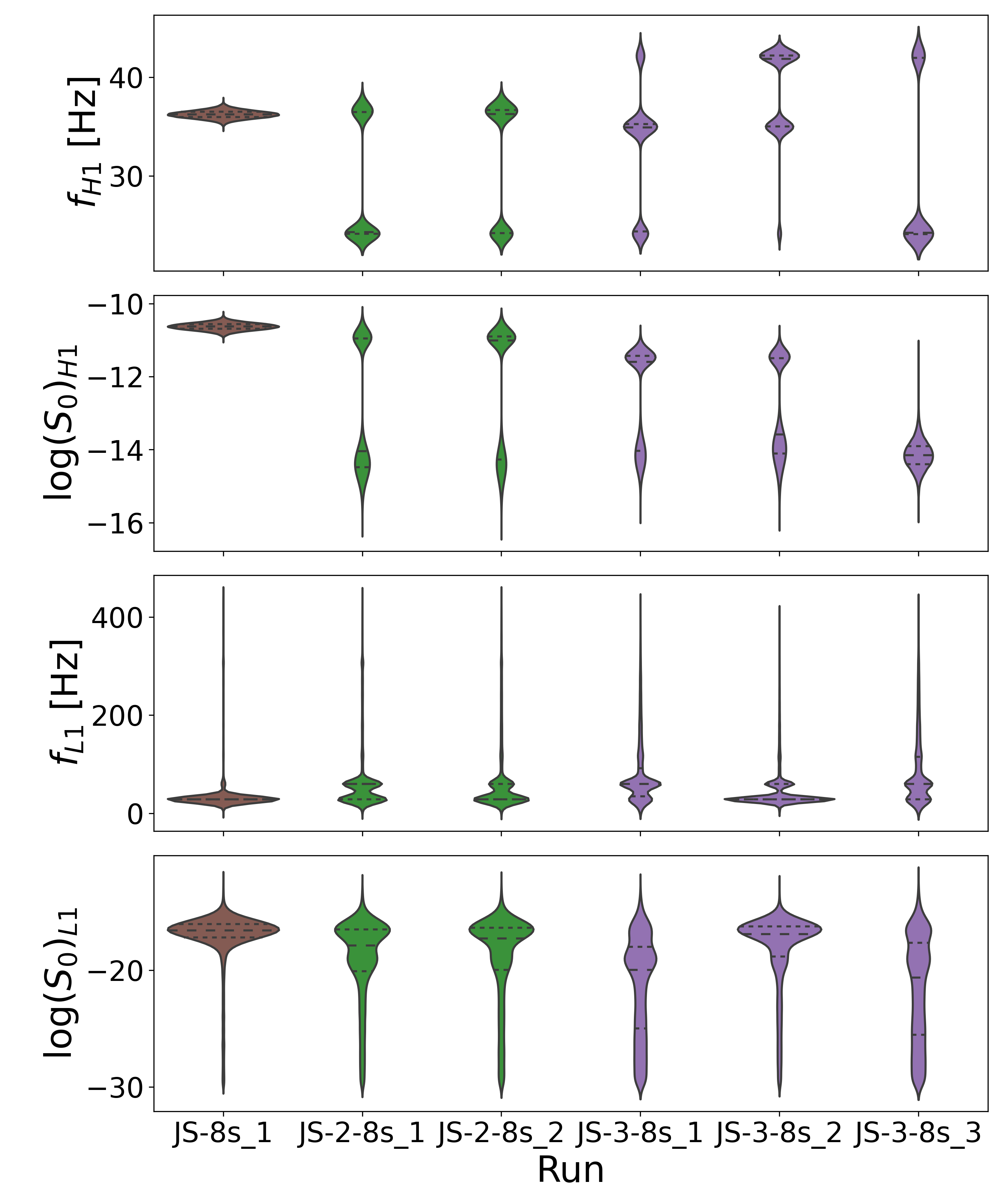} 
\caption{Posterior distributions for the SHO parameters (frequency $f$ and amplitude $\log S_0$) for \GWNineteen. GP-JS, GP-J2S, and GP-J3S denote analyses with one, two, and three SHO terms, respectively; the trailing digit in the $x$-axis labels identifies the individual SHO term. The left panel shows all \qtychecked{4}{\second} runs and the right panel shows all \qtychecked{8}{\second} runs; within each panel, the upper rows show \ac{LHO} and the lower rows show \ac{LLO}. All analyses use the raw data.} \label{fig: GW191109_8s_4s_posteriors_shot_all}
\end{figure*}

We perform a systematic analysis of \GWNineteen using \GPBILBY, with the goal of jointly modelling the gravitational-wave signal and non-Gaussian noise features. 
For both raw and deglitched data, we consider data segments of \qtychecked{4}{\second} and \qtychecked{8}{\second} duration, with the end time fixed to \qtychecked{2}{\second} after merger. 
For each dataset we perform a WL analysis, a GP-J and a GP-JS analysis using the settings employed in the \IMRPhenomXPHMST analysis from \citet{KAGRA:2021vkt}.
The inference results for selected parameters are shown in the left panel of Figure~\ref{fig: GW191109_8s_4s_posteriors}.
Since in the preliminary results, we found support for multiple modes in the GP parameters, when using the raw data frames, we performed additional analyses employing two and three SHO terms with those data frames. 
The inference results for those runs are shown in the right panel of Figure~\ref{fig: GW191109_8s_4s_posteriors}, and the posteriors of the GP parameters for all the runs employing at least one SHO term and raw data are shown in Figure~\ref{fig: GW191109_8s_4s_posteriors_shot_all}.

Including an SHO term in the analysis of the raw data reveals evidence for excess power in both detectors.
The median frequency values found in our analyses are overplotted as red dashed horizontal lines in \cref{fig: GW191109qscan} in \cref{app: spectrograms}.
For the \qtychecked{8}{\second} analysis, the characteristic frequencies inferred for the dominant SHO components are $36.23^{0.65}_{0.63}\,\mathrm{Hz}$ in H1 and $22.35^{+3.63}_{-1.88}\,\mathrm{Hz}$ in L1. 
Comparison with the time-frequency spectograms in Fig.~\ref{fig: GW191109qscan} indicates that the \ac{LLO} feature can be associated with the slow-scattering glitch occurring approximately $\sim 3.2\,\mathrm{s}$ before the trigger.
The interpretation of the \ac{LHO} feature is evident, as the inferred frequencies correspond to the glitch occurring $\sim 4.5\,\mathrm{s}$ before merger, and the same excess power is only partially found in the \qtychecked{4}{\second} analysis.
The power found in the 4s analysis for \ac{LHO} can be related to the residuals of the glitch subtraction.

Overall, the inferred source-parameter posteriors for the \ac{BBH} are largely consistent across configurations, indicating that the parameter estimation is relatively robust to the details of the glitch modelling. 
The inferred posterior on the effective inspiral spin $\chi_{\mathrm{eff}}$ remains qualitatively similar and consistently favours negative values for all analyses.
Interestingly, the non-zero support for non-negative spins seen in the analysis with \BILBY using the deglitched frames for \qtychecked{4}{\second} and \qtychecked{8}{\second} data lengths, vanishes in all the \GPBILBY analyses, especially those including at least one SHO term.
The primary differences arise in the behaviour of the SHO terms themselves. 
In particular, the SHO components recover stronger and more sharply constrained excess power in the \ac{LHO} detector than in \ac{LLO}, consistent with the higher-amplitude glitch observed in H1.
In \ac{LLO}, excess power is also identified, but with smaller amplitudes and broader \qtychecked{90}{\%} credible intervals.

When allowing multiple SHO terms, we find that three SHO components are required to adequately model the glitch structure in the \ac{LHO} data, whereas two SHO terms are sufficient in \ac{LLO}.
Notably, none of the configurations shows evidence for excess power at $\sim \qtychecked{36}{\hertz}$ in \ac{LLO} being captured by the Gaussian-process noise model, consistent with the difficulty of disentangling this feature from the gravitational-wave signal.
 
While the inclusion of explicit glitch modelling does not lead to substantial shifts in the source-parameter posteriors relative to the production analysis, it provides a coherent framework in which the observed non-Gaussian noise features—particularly the stronger glitches in \ac{LHO}—are explicitly identified and marginalised over.


\subsubsection{Comparison with previous glitch-modelling studies}

A detailed investigation of \GWNineteen was recently presented by \citet{Udall:2024ovp}, who analysed this event using a range of waveform approximants and glitch-modelling strategies. In particular, they focused on the excess power at \qtyrange{35}{40}{\hertz} in the \ac{LLO} detector and performed joint signal--glitch inference using both a physically motivated slow-scattering model and a flexible wavelet-based approach implemented in \BAYESWAVE \citep{Cornish:2014kda, Cornish:2020dwh}. Considering several analysis configurations, they found no decisive evidence favouring an interpretation of the excess power as either a glitch or part of the astrophysical signal.

Our results are complementary to those of \citet{Udall:2024ovp}. Rather than attempting to assign the excess power to a specific origin, the \GPBILBY framework marginalises over non-Gaussian noise contributions through a stochastic process model. This allows us to recover inference results consistent with the production analysis while simultaneously accounting for glitch power in the raw data. The agreement between the two approaches strengthens the conclusion that \GWNineteen is robustly characterised by a negative effective spin, independent of the precise treatment of the overlapping noise features.

\subsubsection{Inspiral--merger--ringdown consistency test}

Finally, we perform an inspiral--merger--ringdown (IMR) consistency test by analysing different portions of the data separately using the same \GPBILBY configuration. This enables a direct comparison of the posterior distributions for the final mass and spin inferred from the inspiral and post-inspiral regimes, while consistently modelling non-Gaussian noise.

\begin{figure}[htp!] 
\centering 
\includegraphics[width=\linewidth]{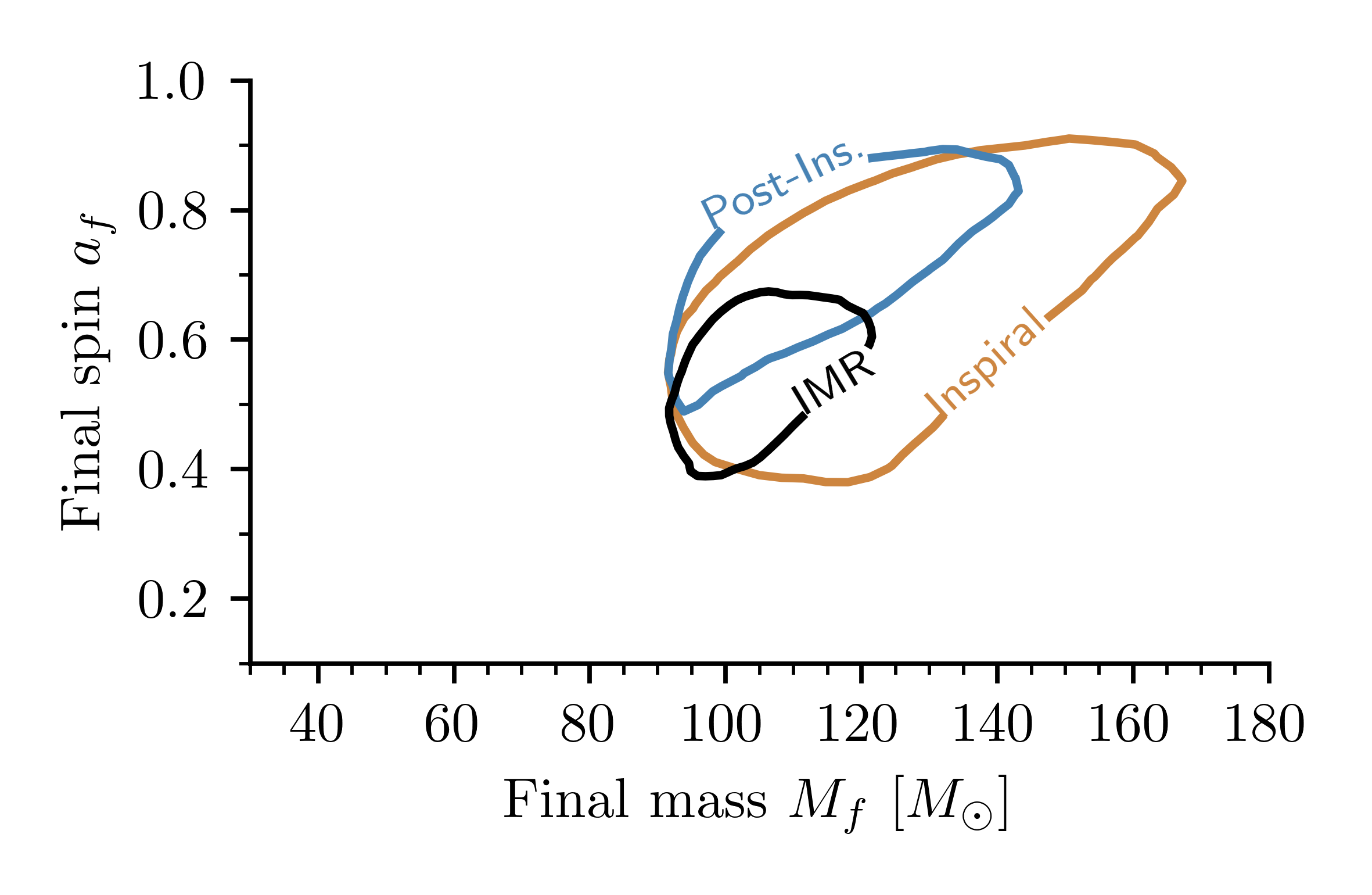} 
\includegraphics[width=\linewidth]{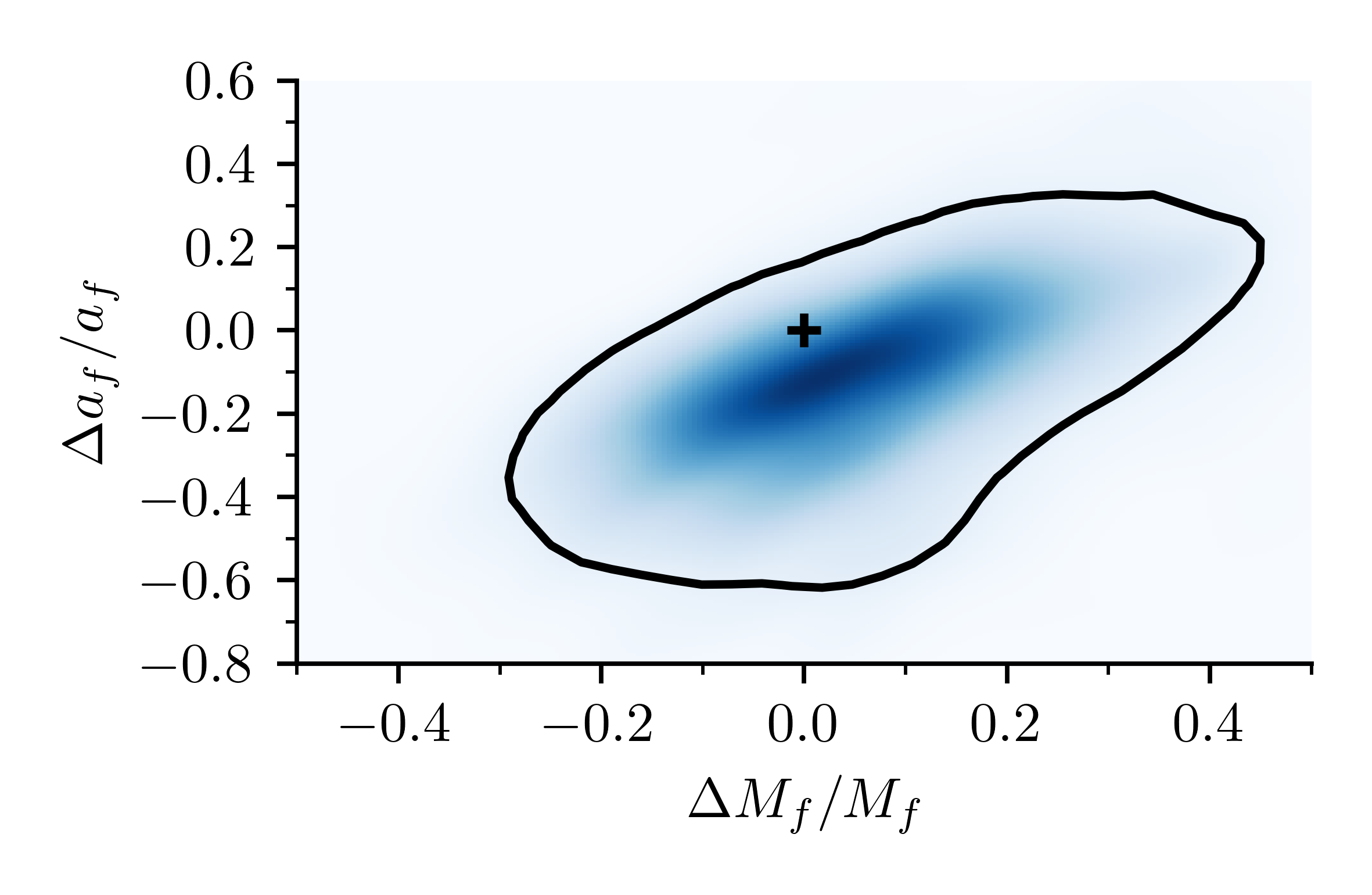} 
\caption{Inspiral merger ringdown test for \GWNineteen. 
The top plot shows the mass and spin estimates of the remnant estimated by using the full signal (in black), using only the inspiral and merger portion (in orange and using just the post-merger part of the signal (in blue).
The bottom plot shows the 90\% credible regions of the
two-dimensional posteriors on $\Delta a_f/a_f$ and $\Delta M_f/M_f$, with (0, 0) being the expected value for general relativity.} 
\label{fig: GW191109_IMR} 
\end{figure}

This analysis is particularly noteworthy because \GWNineteen was excluded from the IMR tests performed in the \ac{O3} tests of general relativity due to concerns about data quality \citep{LIGOScientific:2021sio}. By explicitly modelling glitches alongside the signal, \GPBILBY enables an IMR test for this event despite the presence of strong non-Gaussianities. The results of this test are shown in Fig.~\ref{fig: GW191109_IMR} and demonstrate the potential of Gaussian-process-based methods to extend precision tests of gravity to events previously deemed unsuitable for such analyses.

\subsection{\GWTwentyThreeElevenThirteenLong}
\label{sec: GW231113}
We analyse the event \GWTwentyThreeElevenThirteenLong, first presented in the \GWTCFour catalogue \citep{GWTC-4-results} and observed by the \ac{LHO} and \ac{LLO} detectors.
The signal is relatively weak, with a network \ac{SNR} of just \numchecked{8.6}.
Additionally, the data from \ac{LLO} was found to have data quality issues.
Therefore, deglitched frames were created using \BAYESWAVE and targeting a region from
\qtyrange{70}{120}{\hertz} in the ~\qtychecked{0.2}{\second} window after the merger time.
We therefore use this case study to investigate how well \GPBILBY performs for a typical analysis where we have a single detector impacted by data quality issues and the glitch and signal tracks are reasonably well separated (see~\cref{fig: GW231113qscan} in~\cref{app: spectrograms}).

\begin{figure}[htp!]
    \centering
    \includegraphics[width=\linewidth]{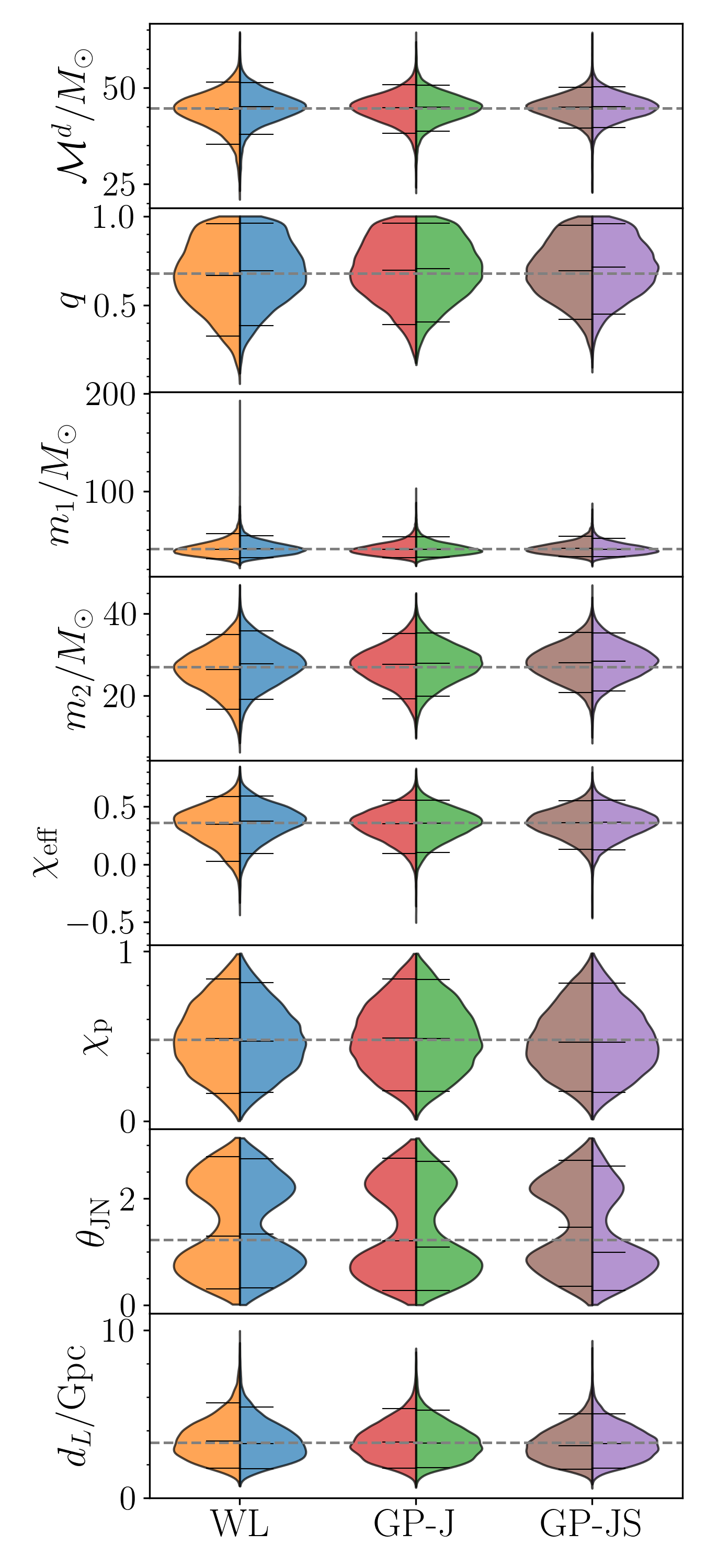}
    \caption{Violin plots of selected \GWTwentyThreeElevenThirteenLong source parameter posteriors for the Whittle likelihood (WL), \GPBILBY with a single jitter term (GP-J), and \GPBILBY with a jitter and SHO term (GP-JS). The left side of each violin shows the posterior obtained from the raw data, while the right side shows the posterior obtained from the deglitched data. The dashed grey line indicates the median value for the \IMRPhenomXPHMST posteriors reported in the discovery paper~\cite{LIGOScientific:2025slb}.}
    \label{fig: GW231113_posteriors}
\end{figure}

We perform a WL analysis, a GP-J and a GP-JS analysis using the settings employed in the \IMRPhenomXPHMST analysis from \citet{LIGOScientific:2025rsn}.
We perform each of these analyses twice, once using the raw data and once using the deglitched data.
In \cref{fig: GW231113_posteriors}, we show the posteriors of selected source parameters using a violin plot comparing the inference from the raw and deglitched frames.
We find overall strong consistency among all the analyses, with the only marked difference being that the WL analysis finds a long tail to high mass in the primary (mirroring the results of the initial analysis \citep{GWTC-4-results}). Meanwhile, the two \GPBILBY analyses are more constraining and do not find support for this long tail.

Examining the posteriors for the \ac{GP} parameters from the GP-JS analysis (plot included in the data release~\cite{emma2026gpbilby_zenodo}), we find evidence of glitch mitigation behavior.
For \ac{LHO}, the posteriors on the frequency are broad and uninformative, while the amplitude and quality factor parameters are peaked at their arbitrary lower bound, indicating no significant glitch power (in agreement with the original analysis).
For \ac{LLO}, the frequency posteriors peak at $\sim$\qtychecked{220}{\hertz} for both the raw and deglitched frames, suggesting \GPBILBY is fitting glitch power at a different frequency than that mitigated in the deglitched frames.
However, the agreement in source parameter estimates (see \cref{fig: GW231113_posteriors}) indicates that this additional power has no meaningful impact on the astrophysical inferences.

\subsection{\GWTwentyThreeElevenTwentyThree}
\label{sec:GW231123}

\GWTwentyThreeElevenTwentyThreeLong (hereafter \GWTwentyThreeElevenTwentyThree) is probably the most massive \ac{BBH} reported to date \citep{LIGOScientific:2025rsn}, with a total mass between \qtyrange{188}{265}{\msun} (though see \citet{Mandel:2025qnh}).
The primary component lies within or above the theorised mass gap, where black holes in the range \qtyrange{60}{130}{\msun} are expected to be rare due to pair-instability processes, while the secondary lies within this gap.
The detection of \GWTwentyThreeElevenTwentyThree was therefore interpreted as evidence that black holes can form through channels other than standard stellar collapse, and that intermediate-mass black holes with masses of order $\sim$\qtychecked{200}{\msun} may be assembled through gravitational-wave-driven hierarchical mergers.
This discovery has prompted renewed discussion of currently accepted theories of stellar evolution and binary black hole formation pathways~\cite{Croon:2025gol}.

The signal was observed by both LIGO detectors, and both data sets are affected by glitches.
In \ac{LHO}, a glitch occurred between \numchecked{1.7} and \qtychecked{1.1}{\second} before the event in the frequency range \qtyrange{15}{30}{\hertz}; this glitch was modelled and subtracted using \BAYESWAVE, and the \ac{LVK} parameter estimates \citep{LIGOScientific:2025rsn, GWTC-4-results} use the resulting deglitched data.
\citet{LIGOScientific:2025rsn} also report additional broadband non-stationary noise in \ac{LHO} around the time of the event.
In the \ac{LLO} detector, a glitch was reported in the frequency range \qtyrange{10}{20}{\hertz} within \qtyrange{3.0}{2.0}{\second} before the event.
However, its time--frequency morphology indicated that it would have no measurable impact on the analysis, and therefore no glitch subtraction was performed.
In \cref{fig: GW231123qscan}, we show time--frequency spectrograms of the raw data for the two LIGO detectors together with the deglitched data for \ac{LHO}.
Because of the short duration of the signal and the presence of non-Gaussian noise transients in the surrounding data, several studies have challenged the current interpretation of the event and proposed alternative explanations of the signal \citep[see, e.g.][]{Passenger:2025acb, Yuan:2025avq, Goyal:2025eqo, Hu:2025lhv, Cuceu:2025fzi, Ray:2025rtt, DeLuca:2025fln}.

We analyse \GWTwentyThreeElevenTwentyThree with \GPBILBY to test whether the astrophysical signal can be cleanly separated from terrestrial disturbances.
We use the same settings as the \IMRPhenomXPHMST analysis of \citet{LIGOScientific:2025rsn} and perform analyses on both the raw and deglitched data.
For each data product, we perform three runs: a WL analysis, a GP-J analysis, and a GP-JS analysis.

\begin{figure}[htp!]
    \centering
    \includegraphics[width=\linewidth]{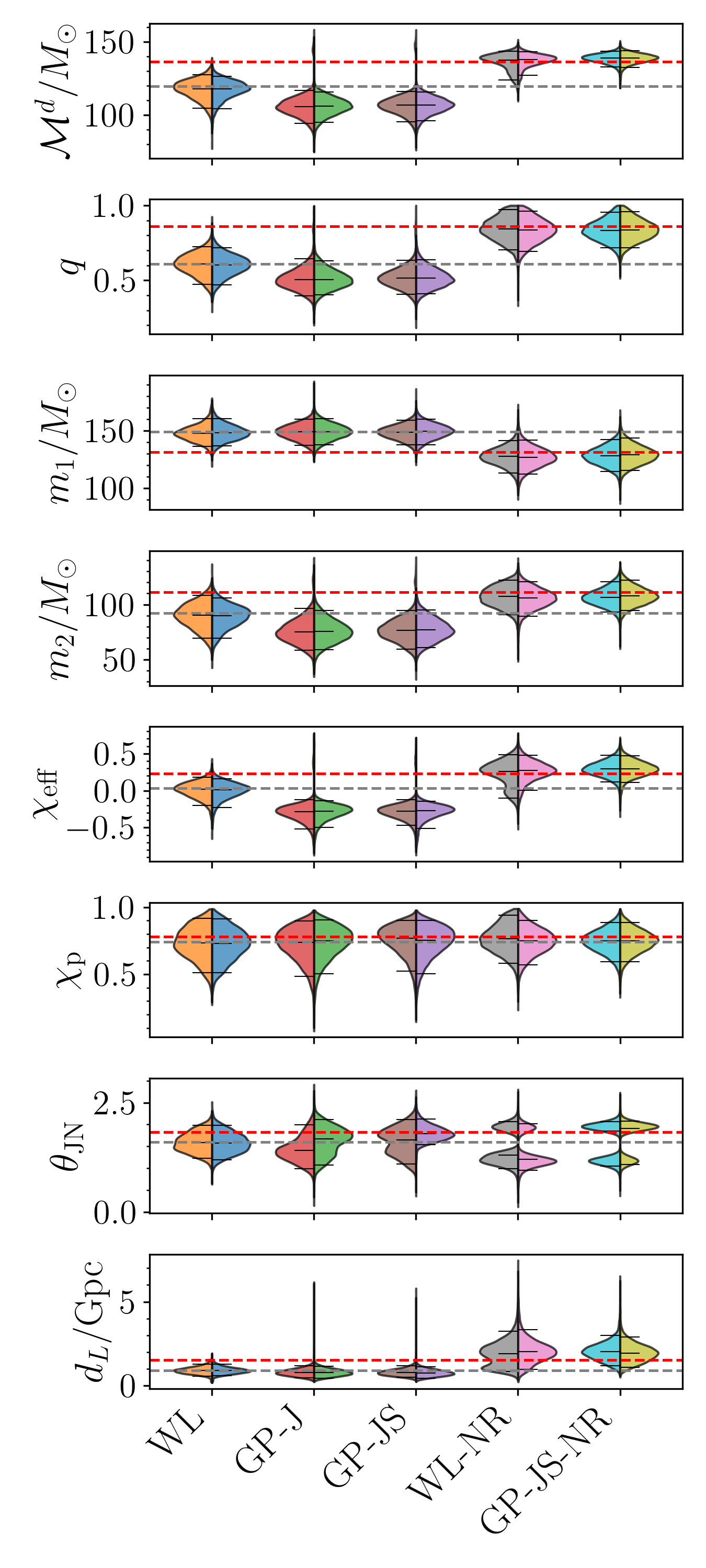}
    \caption{Violin plots of selected \GWTwentyThreeElevenTwentyThree source parameter posteriors for the Whittle likelihood (WL), \GPBILBY with a single jitter term (GP-J), and \GPBILBY with a jitter and SHO term (GP-JS) with \IMRPhenomXPHMST, and for the Whittle likelihood (WL-NR) and \GPBILBY with a jitter and SHO term (GP-JS-NR) with \SURSEVENDQFOUR. The left side of each violin shows the posterior obtained from the deglitched data, while the right side shows the posterior obtained from the raw data. The dashed grey and red line indicates the median value for the \IMRPhenomXPHMST and \SURSEVENDQFOUR posteriors reported in the discovery paper~\cite{LIGOScientific:2025cmm}, respectively.}
    \label{fig: GW231123_posteriors}
\end{figure}

The results are presented in \cref{fig: GW231123_posteriors} using violin plots to compare the inferred source parameters for both deglitched and raw data.
The WL analysis shows good agreement with the original \ac{LVK} results.
By contrast, the GP-J and GP-JS analyses favour a smaller secondary mass, while the primary mass estimates remain consistent.
Even with this reduction, the inference still identifies \GWTwentyThreeElevenTwentyThree as one of the most massive events observed: for the GP-JS analysis on deglitched data, we infer a total mass of ${227}_{-23}^{+22}$~\Msun.
In addition, the \GPBILBY analyses show support for a negative effective spin parameter.
This suggests that at least one component has significant spin misalignment with the orbital angular momentum, which is more consistent with dynamical assembly than with isolated binary evolution.

To further investigate these results, we perform an additional analysis using the \SURSEVENDQFOUR waveform model with the settings employed in~\citet{LIGOScientific:2025rsn}.
We analyse both the raw and deglitched data using the Whittle likelihood and \GPBILBY with a jtter term and a single SHO term.
The results, shown in the two rightmost violins in \cref{fig: GW231123_posteriors}, indicate that in this case no significant difference is recovered between the \BILBY and \GPBILBY analyses.

\begin{figure*}[htp!]
    \centering
    \includegraphics[width=\linewidth]{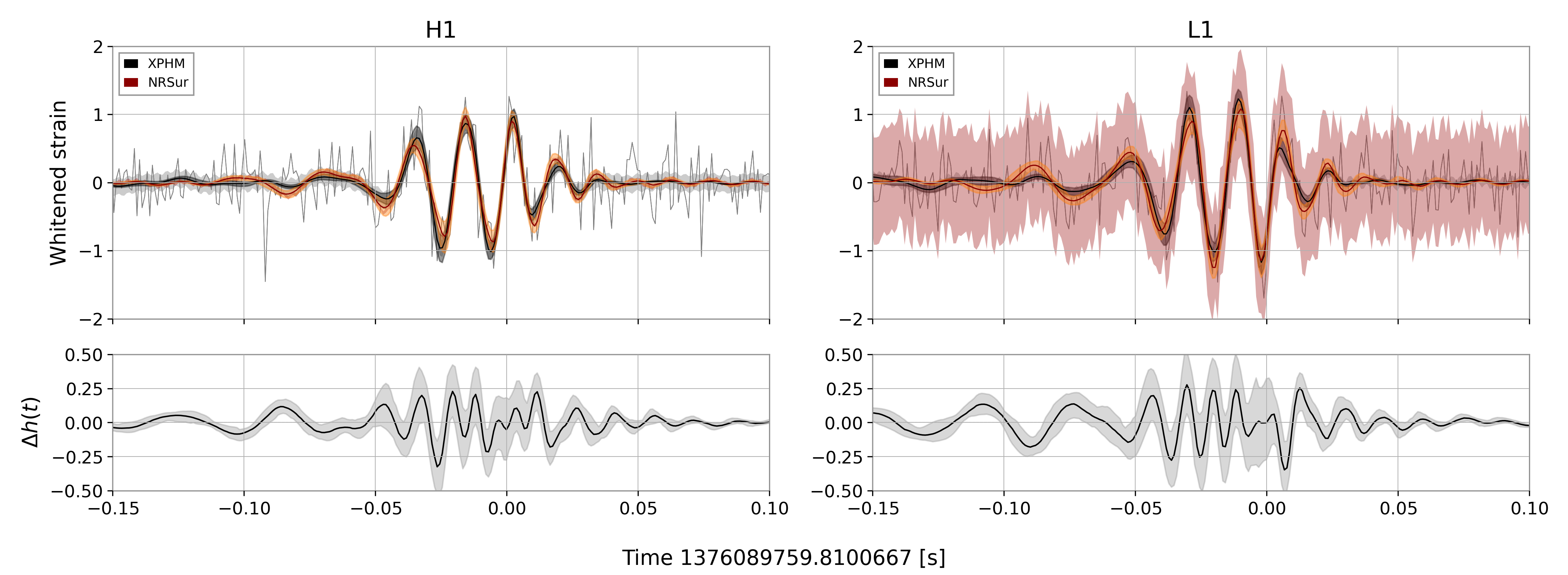}
    \caption{Posterior predictive plots for the event \GWTwentyThreeElevenTwentyThree for the \ac{LHO} and \ac{LLO} detectors, shown in the left and right columns, respectively.
The upper panels display the reconstructed strain obtained from a \GPBILBY analysis using one SHO term with the \IMRPhenomXPHMST waveform and the \SURSEVENDQFOUR waveform, for the Hanford and Livingston detectors on the left and right, respectively.
The residuals between the two waveform reconstructions are shown in the bottom panels.
Differences in the whitened strain relative to Fig.~1 of~\citet{LIGOScientific:2025rsn} arise from the use of different data-conditioning procedures.}
    \label{fig: GW231123_predictive plots}
\end{figure*}

A clearer picture emerges when directly comparing the waveform reconstructions obtained with the two waveform families, shown in \cref{fig: GW231123_predictive plots}.
Although both models provide acceptable fits to the data at the level of the overall signal morphology, the reconstructed waveforms exhibit systematic differences.
In particular, the NRSur-based reconstruction predicts noticeably different amplitudes and phases around the merger peaks, while smaller but coherent deviations are also visible during the late inspiral.
These differences are most apparent in the \ac{LLO} detector owing to its higher \ac{SNR} ratio; however, when considered relative to the signal amplitude, the residual structures are comparable in both detectors.
The bottom panels, showing the residuals between the two reconstructions, indicate that these deviations are coherent features concentrated primarily in the highest-frequency portion of the signal rather than random fluctuations.

The presence of such coherent residuals provides a natural explanation for the behaviour observed in the parameter-estimation results.
In particular, the inclusion of Gaussian-process terms leads to shifts in the inferred parameters for the \IMRPhenomXPHMST analyses, while the corresponding \SURSEVENDQFOUR results remain stable.
This suggests that, for \IMRPhenomXPHMST, part of the mismatch between waveform and data is absorbed by the GP model, whereas for \SURSEVENDQFOUR the waveform reconstruction leaves comparatively little coherent structure to be captured by the GP.
The improved agreement between waveform and data in \SURSEVENDQFOUR allows for the inference to remain consistent between the WL and GP analyses.

This interpretation highlights that the behaviour of the GP-enhanced likelihood can be intrinsically coupled to waveform accuracy.
If the waveform model captures the signal morphology well, the GP terms remain subdominant and primarily account for genuine noise features.
If instead the waveform leaves coherent residuals, the GP can absorb part of this structure, effectively redistributing power between the signal and noise models.
The resulting posterior shifts therefore provide indirect evidence for modelling systematics.

These findings should be interpreted in the context of the analysis of \citet{Bini:2026kwz}, who argued that waveform systematics and Gaussian-noise effects can account for the observed behaviour of \GWTwentyThreeElevenTwentyThree and that the \SURSEVENDQFOUR-based interpretation remains robust.
While our results are broadly consistent with the overall astrophysical picture, they highlight an aspect that becomes apparent when using a flexible likelihood model.
Although Bini et al.\ note that the waveform models are highly similar (as quantified by differences in the whitened signal), our waveform reconstructions reveal coherent and visibly significant discrepancies in key parts of the signal, both in the merger cycles and in the preceding portion of the waveform.
These differences are large compared to the local signal structure and lead to systematic shifts in the inferred source parameters between the two waveform families, even though the qualitative astrophysical interpretation remains unchanged.

Such localised discrepancies leave structured residuals after subtraction that can be captured with a GP-enhanced likelihood.
From this perspective, the GP analysis acts as a diagnostic of waveform mismodelling: waveform differences that may appear subdominant in global comparisons can become clearly visible through their residual structure and their effect on inference.

\GWTwentyThreeElevenTwentyThree therefore provides a useful case study illustrating how waveform systematics, noise modelling, and parameter inference are tightly coupled within the analysis.
Within a GP-enhanced likelihood framework, coherent residuals arising from waveform differences can be partially absorbed by the noise model, translating modelling discrepancies into shifts in the inferred astrophysical parameters.
The comparison between \IMRPhenomXPHMST and \SURSEVENDQFOUR shows that this behaviour depends sensitively on waveform accuracy.
When coherent waveform residuals are present, GP terms can absorb part of the signal structure and modify the balance between signal and noise contributions.
In contrast, when the waveform provides a closer representation of the data, the GP contribution remains subdominant and the inferred parameters are stable across likelihood choices.
In this sense, the GP analysis acts as a probe of residual structure that may not be evident from global waveform comparisons alone.
These results motivate treating waveform modelling and likelihood flexibility as jointly interacting components when interpreting individual high-mass events.

\subsection{\GWTwentyThreeSixLong}
\label{sec: GW230630}
\GWTwentyThreeSixLong is a candidate gravitational-wave signal identified in \ac{O4a} by the \GSTLAL pipeline \citep{Messick:2016aqy,Sachdev:2019vvd,Tsukada:2023edh,Sakon:2022ibh,Joshi:2025nty}, with a \ac{FAR} of \numchecked{0.47} per year and a \pastro of \numchecked{0.88} \citep{GWTC-4-results}.
However, offline follow-up \citep[as described in Ref.][]{LIGO:2024kkz} identified this event as likely of instrumental origin, due to excess power in both detectors, consistent with scattered light \citep{Ottaway:2012oce}.
In \cref{fig: GW230630_spectrograms}, we reproduce the time-frequency spectrograms first reported in \citet{GWTC-4-results}, but with an extended time span.

To further investigate \GWTwentyThreeSixLong, we perform WL and GP-JS analyses of the event using the \IMRPhenomXPHM waveform.
We know that the signal must have a reasonably good match with the \ac{BBH} model, since it was picked up by a modelled search pipeline.
However, our motivation is to test if, after fitting the \ac{BBH} signal, there is any residual power modelled by the \ac{GP}.
This would be indicative of a non-\ac{BBH} source, e.g. an unfortunate coincidence between two otherwise independent glitches.

Since parameter estimation was not performed in the original analyses, we set up new analyses, but followed the same approach as applied to other \GWTCFour candidates \citep{GWTC-4-methods}.

\begin{figure}[htp!]
    \centering
    \includegraphics[width=\linewidth]{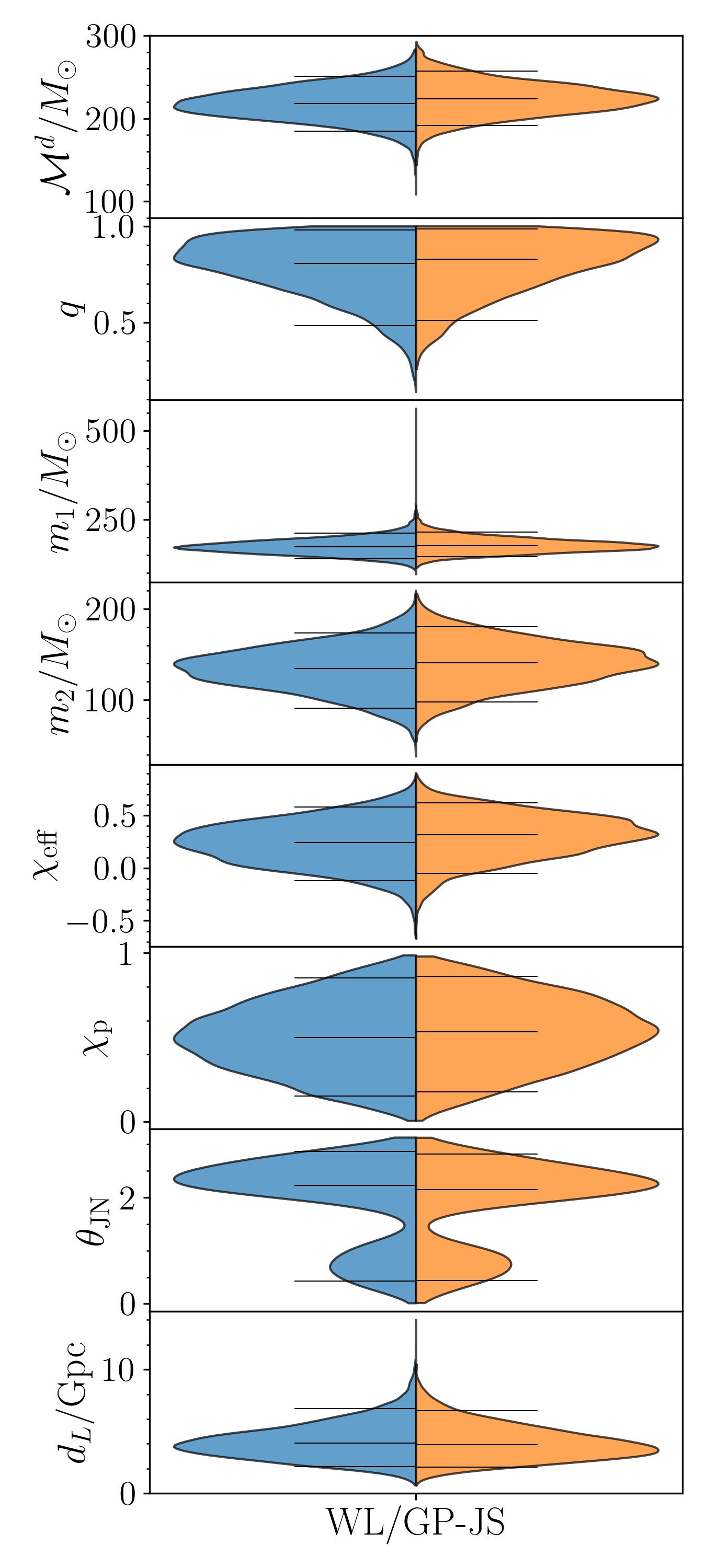}
    \caption{Violin plots of selected \GWTwentyThreeSixLong source parameter posteriors for the Whittle likelihood (WL) and \GPBILBY with a jitter and SHO term (GP-JS).}
    \label{fig: GW230630_violins}
\end{figure}

In \cref{fig: GW230630_violins}, we present the inferred source parameters.
The comparison demonstrates good agreement between the two analyses, with shifts in the median below the uncertainty in the posterior distribution itself.
We infer that, if real, \GWTwentyThreeSixLong is an exceptionally high-mass event: the primary source mass is inferred to be ${173}_{-33}^{+39}$~\Msun by the WL analysis.
This is significantly larger than \GWTwentyThreeElevenTwentyThree, which has a 
primary mass of $137^{+23}_{-18}$ \citep{LIGOScientific:2025rsn}.
Therefore, this would make the primary of \GWTwentyThreeSixLong one of the largest black holes observed to date.
Moreover, the secondary of \GWTwentyThreeSixLong is also massive, at ${135}_{-44}^{+39}$~\Msun.
Combined, the total mass would be ${306}_{-53}^{+64}$, again exceeding that of \GWTwentyThreeElevenTwentyThree, the current record holder.

Meanwhile, for the inferred \ac{GP} parameters (the plot is provided in the data release~\cite{emma2026gpbilby_zenodo}), the frequency posteriors are uninformative and the amplitude posterior peaks at the arbitrary lower bound. 
This indicates that, within the sensitivity of our model, we do not identify additional residual power beyond that captured by the \ac{BBH} waveform.
We emphasize, however, that this result should be interpreted with caution.
In particular, the absence of significant residual power in the \GPBILBY analysis does not imply that the data are consistent with purely Gaussian noise or that the event is astrophysical in origin. 
Our method assumes a \ac{BBH} signal model and tests for additional structured deviations, but it is not designed to distinguish between astrophysical signals and coincident or incoherent noise artefacts.
This is especially relevant in light of the \GWTCFour analysis, which identified excess power in both detectors inconsistent with a \ac{CBC} signal and attributed the candidate to instrumental noise. 
The apparent consistency of the data with a \ac{BBH} model in our analysis may therefore reflect the limited duration and low signal-to-noise ratio of the candidate, which make it difficult to robustly distinguish between competing interpretations.
We therefore conclude that, while \GWTwentyThreeSixLong can be fit by a \ac{BBH} waveform without requiring additional glitch modelling in this framework, our analysis does not provide evidence in favour of an astrophysical origin.


\section{Discussion and Conclusion}
\label{sec: discussion}




In this work, we have presented a set of case studies using \GPBILBY to model observed gravitational-wave signals in both Gaussian and non-Gaussian noise environments.
By jointly modelling the astrophysical signal and transient noise artefacts with a time-domain Gaussian process, we demonstrate a flexible alternative to traditional glitch-subtraction approaches.

For events consistent with Gaussian noise, such as \GWFifteen and \GWSeventeen, \GPBILBY yields posterior distributions in good agreement with the standard Whittle likelihood.
In \GWSeventeen, the GP analysis produces tighter credible intervals and reduces the weight of secondary modes in the inclination-angle posterior.
In \GWFifteen, \GPBILBY also identifies narrowband instrumental features, with the SHO terms recovering the \qtychecked{60}{\hertz} power line, while leaving the astrophysical inference effectively unchanged.

We then investigated glitch-contaminated events.
For \GWNineteen, the evidence for a negative effective inspiral spin remains robust when analysing raw data with \GPBILBY, showing that this conclusion is not driven by transient noise.
This event additionally enabled an inspiral--merger--ringdown consistency test that was previously excluded owing to data-quality concerns.
Similarly, for \GWTwentyThreeElevenThirteen, \GPBILBY proved more constraining than standard analyses by identifying glitch power without introducing significant bias in the inferred astrophysical parameters.

The case study of \GWTwentyThreeElevenTwentyThree provides a particularly informative example of how GP-enhanced likelihoods can probe waveform performance in high-mass events.
When using \IMRPhenomXPHM, \GPBILBY identifies coherent residual structure that leads to measurable shifts in the inferred mass and spin parameters, indicating that non-negligible signal power remains after waveform subtraction.
By contrast, analyses based on the \SURSEVENDQFOUR waveform remain stable across likelihood choices and show little excess residual power, suggesting that this model captures the signal morphology more completely.
This comparison highlights the diagnostic role of the GP framework: rather than introducing additional bias, the GP terms reveal residual coherent structure when waveform mismatches are present and remain subdominant when the waveform accurately describes the data.
Our results, therefore, reinforce the robustness of the NRSur-based interpretation of \GWTwentyThreeElevenTwentyThree, while showing that flexible likelihood models can make subtle waveform differences directly visible through their impact on inference.
Consistent with the analysis of \citet{Bini:2026kwz}, we find that the overall astrophysical interpretation remains unchanged, but that GP-based analyses provide a complementary way to identify and quantify waveform-related systematics.

A related behaviour is observed in our analysis of \GWTwentyThreeEight, another high-SNR \ac{BBH}, where the inclusion of an SHO term leads to different parameter inference results and to non-zero residual structure and a small amount of power assigned to the GP component, particularly near the peaks of the time-domain signal.
In this case, the effect is likely driven by the very high \ac{SNR}, which renders subtle waveform inaccuracies directly visible in the residuals, rather than by challenges associated with the underlying parameter space.
Together with \GWTwentyThreeElevenTwentyThree, this illustrates that GP-enhanced analyses can act as sensitive probes of small waveform mismatches, both in very high-SNR events and in regions of parameter space in which waveform systematics are expected to be significant.

Finally, we analysed \GWTwentyThreeSixLong, an event excluded from \GWTCFour due to suspected instrumental contamination. 
We find that the data can be fit by a \ac{BBH} waveform without requiring additional structured residuals in the \GPBILBY analysis. 
However, we emphasize that this does not provide evidence for an astrophysical origin, and is not in tension with the \GWTCFour classification. 
The limited duration and low signal-to-noise ratio of the candidate make it difficult to distinguish between a true signal and noise artefacts within this framework.
If astrophysical, it would correspond to one of the highest-mass binaries observed to date, motivating further investigation by the broader community.

Overall, our results demonstrate that \GPBILBY is a powerful diagnostic tool for studying exceptional gravitational-wave events and for separating non-Gaussian noise features from astrophysical signals.
At the same time, the \GWTwentyThreeElevenTwentyThree and \GWTwentyThreeEight analyses show that GP-based likelihoods implicitly assume waveform accuracy: when this assumption fails, residual signal power can be absorbed by the noise model.
In addition, subthreshold or weakly coherent glitches may not be fully captured by the current GP parameterisation, and could therefore remain partially hidden within the inferred noise model.
This highlights the importance of continued development of GP kernel choices and more systematic studies of how kernel flexibility, waveform systematics, and noise transients interact in realistic detector data.
Waveform modelling and flexible noise descriptions must therefore be considered jointly when interpreting high-mass or otherwise challenging gravitational-wave events.

\section*{Data Availability}
The notebooks and scripts to reproduce the results, as well as the result files to reproduce the plots of this study, are publicly available on the Zenodo repository~\cite{emma2026gpbilby_zenodo} and can be used with the \texttt{gpbilby} package available on pypi~\cite{emma2026gpbilby}.

\section*{Acknowledgements}
We would like to thank Ling (Lilli) Sun for the discussion that inspired our treatment of calibration errors in \GPBILBY.
We are also grateful to Michael Williams and Sylvia Biscoveanu for their useful comments and suggestions on the analysis of GW231123. 
This work is supported by the Science and Technology Facilities Council (STFC) grant UKRI2488.

The authors are grateful for computational resources provided by the LIGO Laboratory and supported by National Science Foundation Grants PHY-0757058 and PHY-0823459.
The authors are also grateful for computational resources provided by Cardiff University and supported by STFC grants ST/I006285/1 and ST/V005618/1.
This research has made use of data or software obtained from the Gravitational Wave Open Science Center (gwosc.org), a service of the LIGO Scientific Collaboration, the Virgo Collaboration, and KAGRA. 
This material is based upon work supported by NSF's LIGO Laboratory which is a major facility fully funded by the National Science Foundation, as well as the Science and Technology Facilities Council (STFC) of the United Kingdom, the Max-Planck-Society (MPS), and the State of Niedersachsen/Germany for support of the construction of Advanced LIGO and construction and operation of the GEO600 detector. 
Additional support for Advanced LIGO was provided by the Australian Research Council. 
Virgo is funded, through the European Gravitational Observatory (EGO), by the French Centre National de Recherche Scientifique (CNRS), the Italian Istituto Nazionale di Fisica Nucleare (INFN) and the Dutch Nikhef, with contributions by institutions from Belgium, Germany, Greece, Hungary, Ireland, Japan, Monaco, Poland, Portugal, Spain. KAGRA is supported by the Ministry of Education, Culture, Sports, Science and Technology (MEXT), Japan Society for the Promotion of Science (JSPS) in Japan; the National Research Foundation (NRF) and Ministry of Science and ICT (MSIT) in Korea; Academia Sinica (AS) and National Science and Technology Council (NSTC) in Taiwan.

We utilise the \NUMPY \citep{harris2020numpy} and \SCIPY library \citep{2020SciPy-NMeth} for data processing and analysis, and we also use the \MATPLOTLIB \citep{Hunter:2007ouj} library for visualisation.

\bibliography{references}

\appendix
\onecolumngrid
\section{Spectrograms of gravitational-wave events}~\label{app: spectrograms}

\begin{figure*}[hbp!] 
\centering 
\includegraphics[width=\textwidth]{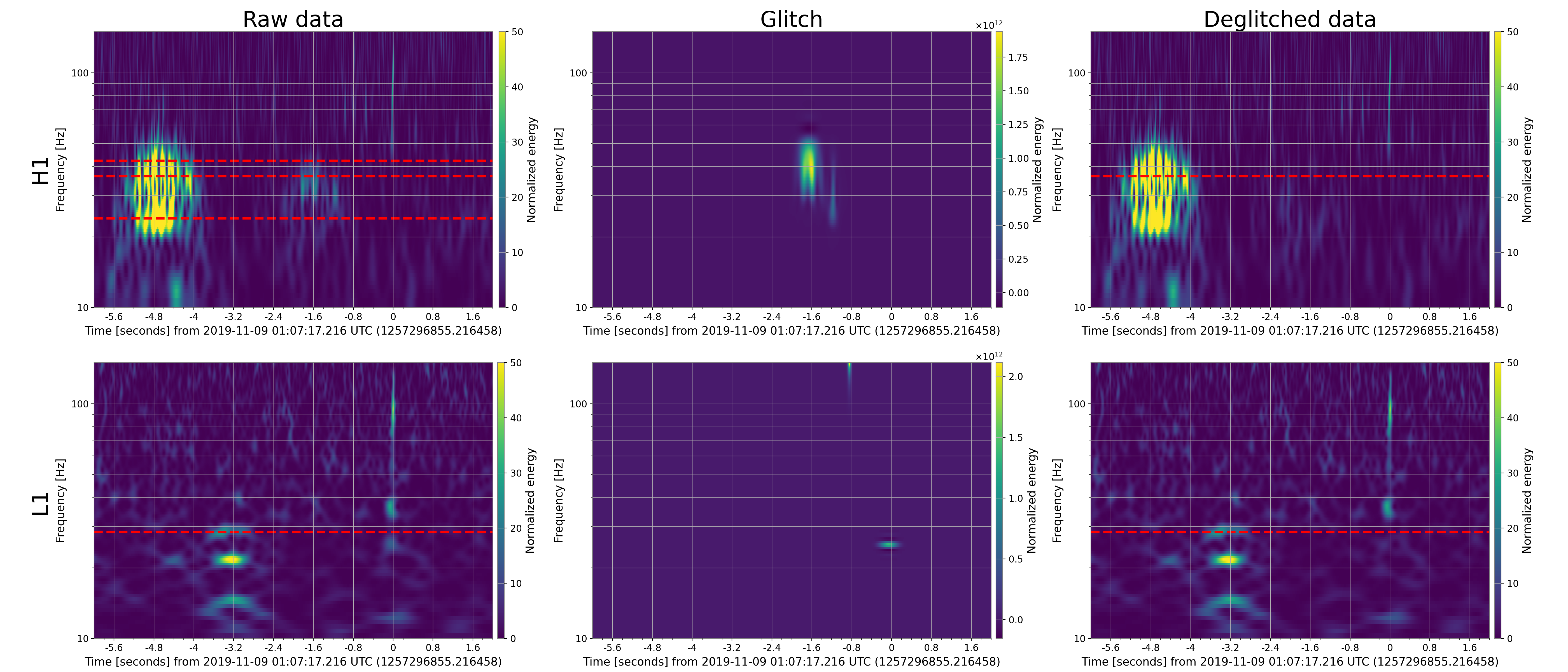} 
\caption{Time-frequency spectrograms of \qtychecked{8}{\second} of data surrounding \GWNineteenLong for the \ac{LHO} detector (top) and the \ac{LLO} detector (bottom). For each detector, the panels show from left to right the raw data, the reconstructed glitch, and the glitch-subtracted data.}
\label{fig: GW191109qscan} 
\end{figure*}




\begin{figure*}[htp!]
    \centering
    \includegraphics[width=\textwidth]{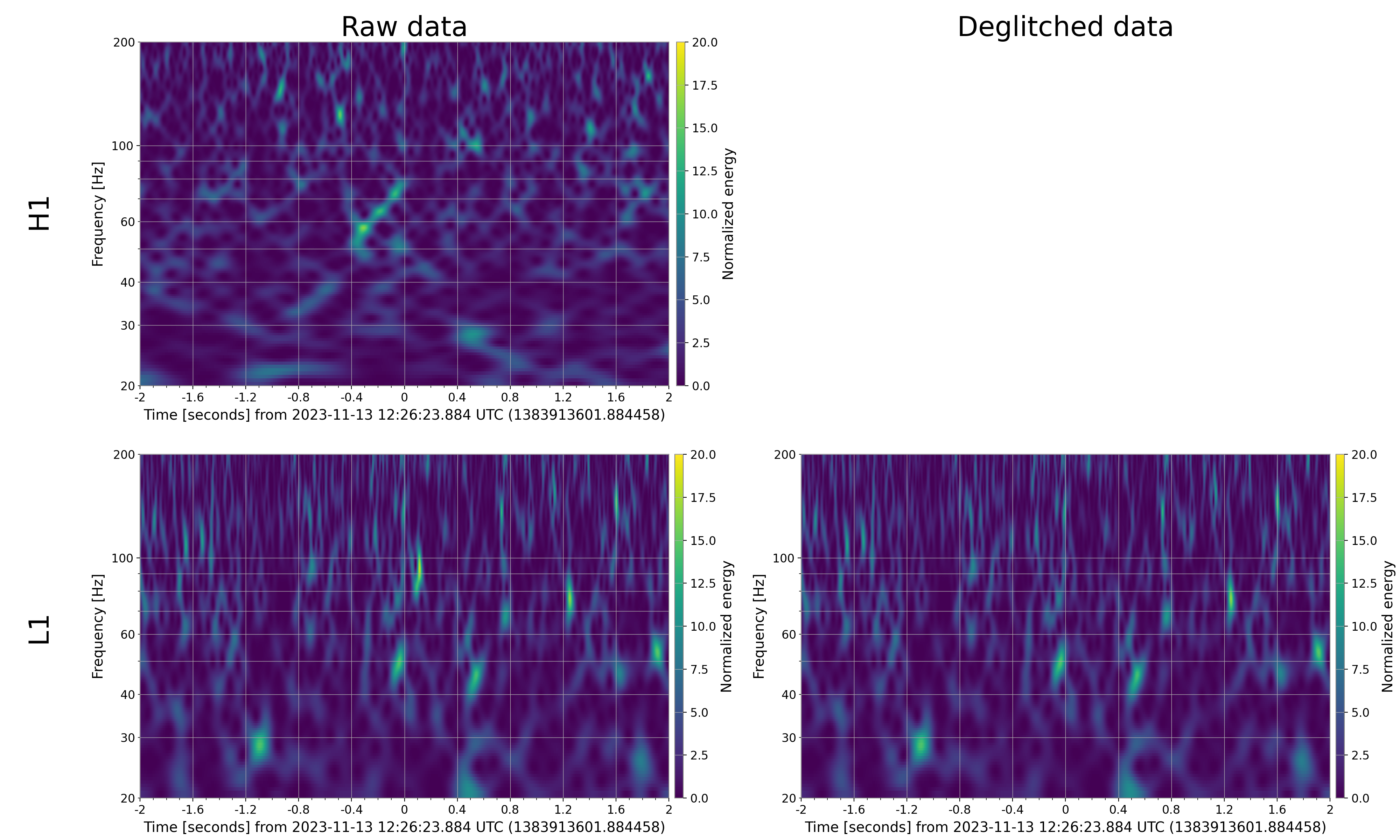}
    \caption{Time-frequency spectrograms of the data surrounding \GWTwentyThreeElevenThirteenLong. The left panels show the raw data for the \ac{LHO} and \ac{LLO} detectors, upper and lower panels, respectively. The right panels show the corresponding deglitched data; no deglitched frames were produced for \ac{LHO}, so the raw data are shown for both columns. The glitch subtracted from the \ac{LLO} data had a frequency range of $\sim$80--100\,Hz and occurred $\sim$0.1\,s after the trigger time. We include two seconds of data before and two seconds of data after the trigger time.}
    \label{fig: GW231113qscan}
\end{figure*}

\begin{figure*}[htp!]
    \centering
    \includegraphics[width=\textwidth]{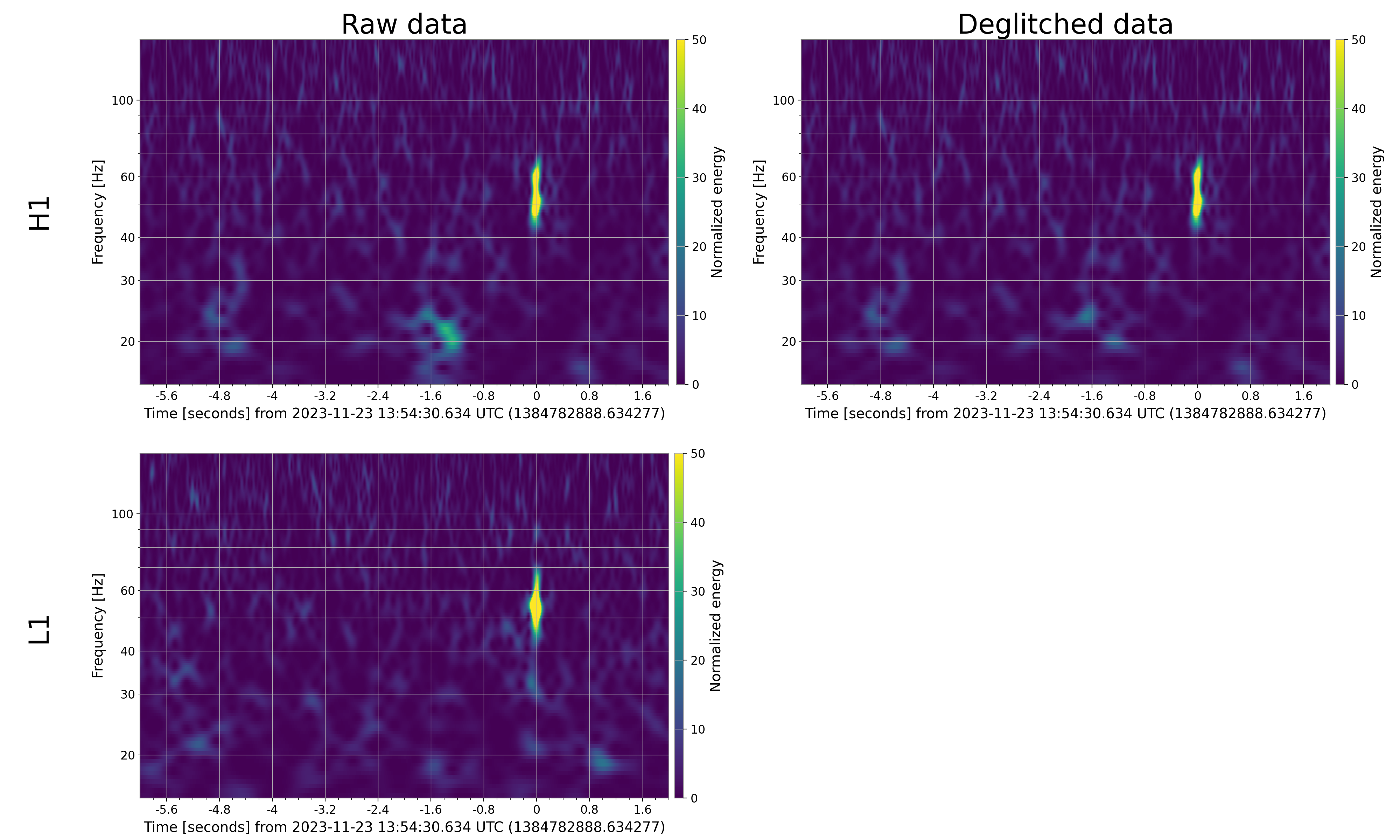}
    \caption{Time-frequency spectrograms of the data surrounding \GWTwentyThreeElevenTwentyThree. The left panels show the raw data for the \ac{LHO} and \ac{LLO} detectors, upper and lower panels, respectively. The right panel shows the deglitched \ac{LHO} data; no deglitched frames were produced for \ac{LLO}. The glitch subtracted from the \ac{LHO} data had a frequency range of $\sim$15--25\,Hz and occurred $\sim$1.5\,s before the trigger time. We include six seconds of data before and two seconds of data after the trigger time.}
    \label{fig: GW231123qscan}
\end{figure*}



\begin{figure*}
    \centering
    \includegraphics[width=\linewidth]{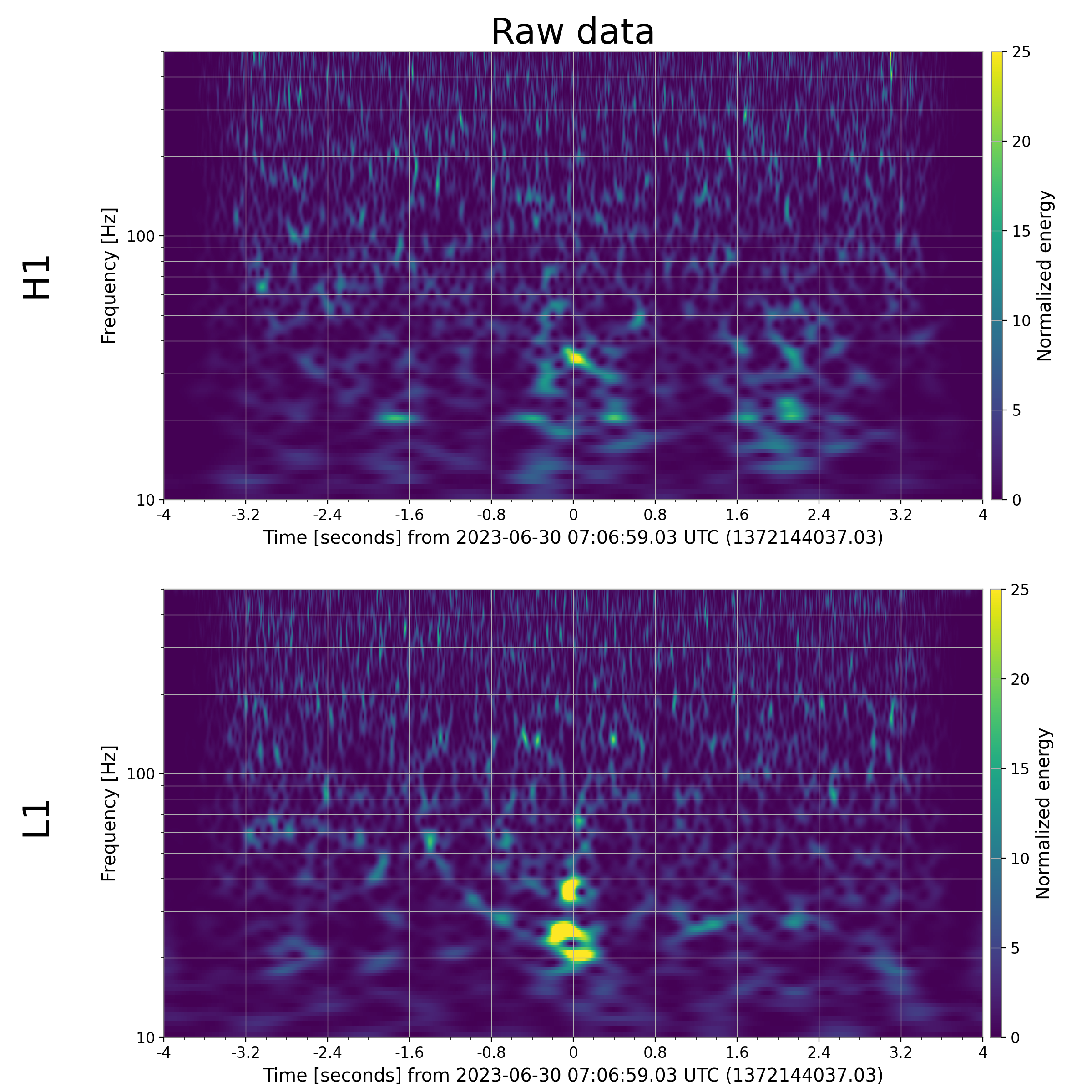}
    \caption{Time-frequency spectrogram of \qtychecked{8}{\second} of data surrounding \GWTwentyThreeSixLong for the \ac{LHO} and \ac{LLO} detectors.}
    \label{fig: GW230630_spectrograms}
\end{figure*}

\end{document}